\documentclass[a4paper,11pt,JHEP]{article}

\usepackage{amssymb}
\usepackage{amsmath}
\usepackage{amscd}
\usepackage{latexsym}
\usepackage{epsfig}
\usepackage{color}
\usepackage{fancyhdr}
\usepackage[bf,footnotesize]{caption}
\usepackage{graphicx}
\usepackage{booktabs}
\usepackage{a4wide}
\usepackage{bbm} 
\usepackage{mathrsfs}
\usepackage{dsfont}
\usepackage{wasysym}
\usepackage{array}
\usepackage[all]{xy}

\usepackage{slashed}
\usepackage[numbers,sort&compress]{natbib}

\usepackage[colorlinks=true,a4paper=true,backref=false, linktocpage=true,citecolor=blue,urlcolor=blue,linkcolor=blue,pdfpagemode=UseOutlines]{hyperref}

\hypersetup{%
  bookmarksnumbered=true,
  pdftitle = {Supersymmetric quantum mechanics on the lattice: I. Loop formulation},
  pdfsubject = {},
  pdfauthor = {U.~Wenger},
  pdfkeywords = {}
  linkcolor=blue,citecolor=blue,urlcolor=blue
}

\graphicspath{{Figures_LoopFormulation/}}

\def\det{\mathop{\rm det}}              
\def\Dslash{\hbox{D}\kern-0.6em\raise0.15ex\hbox{/}} 

\def\bi{\begin{itemize}}
\def\ei{\end{itemize}}
\def\be{\begin{equation}}
\def\ee{\end{equation}}

\newcommand{\psibar}{\overline{\psi}}
\newcommand{\e}{\mathrm{e}}

\newcommand{\Tr}{\mathrm{Tr}}

\renewcommand{\O}{\mathcal{O}}

\newcommand{\G}{\mathcal{G}}
\newcommand{\Ncal}{\mathcal{N}}

\newcommand{\Z}{\mathcal{Z}}
\newcommand{\D}{\mathcal{D}}
\newcommand{\C}{\mathcal{C}}

\date{\empty}

\begin{document}
\title{Supersymmetric quantum mechanics on the lattice:\\
I. Loop formulation}

\author{David Baumgartner and Urs Wenger\vspace{0.5cm}
\\
Albert Einstein Center for Fundamental Physics,\\
Institute for Theoretical Physics, University of Bern, \\
Sidlerstrasse 5, CH--3012 Bern, Switzerland\vspace{0.5cm}
\\
}

\maketitle

\begin{abstract}
  Simulations of supersymmetric field theories on the lattice with
  (spontaneously) broken supersymmetry suffer from a fermion sign
  problem related to the vanishing of the Witten index. We propose a
  novel approach which solves this problem in low dimensions by
  formulating the path integral on the lattice in terms of fermion
  loops.  For $\Ncal = 2$ supersymmetric quantum mechanics the loop
  formulation becomes particularly simple and in this paper - the
  first in a series of three - we discuss in detail the reformulation
  of this model in terms of fermionic and bosonic bonds for various
  lattice discretisations including one which is $Q$-exact.
\end{abstract}

\section{Introduction}
Independently of whether or not supersymmetry is realised in high energy
particle physics, supersymmetric quantum field theories remain to be
interesting and fascinating on their own.  One intriguing feature of
supersymmetric theories is for example the emergence of a Goldstino
mode when the supersymmetry is broken, or the appearance of mass
degenerate multiplets of fermionic and bosonic particles if the ground
state of the theory is invariant under the supersymmetry
transformation.  In nature though, such degeneracies among elementary
particles have so far not been observed, and as a consequence the
supersymmetry must be spontaneously broken at some scale
\cite{O'Raifeartaigh1975331} if supersymmetry is indeed a true
symmetry of nature. In fact, the spontaneous breaking of supersymmetry
is a generic phenomenon which is relevant for many physical systems
beyond particle physics and quantum field theories.  The question of
spontaneous supersymmetry breaking, however, cannot be addressed in
perturbation theory and nonperturbative methods are therefore
desirable and even crucial.  In the past, numerical simulations of
quantum field theories on Euclidean lattices have proven to be a very
successful tool for studying nonperturbative phenomena. Consequently,
a lot of effort has been put into the lattice formulation of
supersymmetric field theories, e.g.~\cite{Montvay:1996pz,
  Kaplan:2002wv, Feo:2004kx, Giedt:2006pd, Takimi:2007nn,
  Damgaard:2007eh, Catterall:2007kn}, see \cite{Catterall:2009it} for
a comprehensive review.
Finding an appropriate formulation, however, turns out to be far from
trivial due to the explicit breaking of symmetries in connection with
the discretisation. The Poincar\'e group for example is broken down to
the subgroups of discrete rotations and finite translations by
multiples of the lattice spacing. Since supersymmetry is an extension
of the Poincar\'e algebra, a complete realisation of the continuum
supersymmetry algebra on the lattice is therefore not possible. For
lattice regularised theories which are composed of local lattice
operators, however, the remnant subgroups guarantee that the
Poincar\'e symmetry is fully restored in the continuum. Unfortunately,
in contrast to the Poincar\'e symmetry, for supersymmetry there is in
general no subgroup left on the lattice which could play the role of
the discrete subgroups above. It is therefore a priori not clear at
all how a lattice formulation can be found for which supersymmetry is
restored in the continuum \cite{Bartels:1983}, a problem which can
eventually be traced back to the failure of the Leibniz rule on the
lattice \cite{Dondi:1977, Fujikawa:2002ic, Bruckmann:2006ub}.

Apart from the explicit breaking of supersymmetry by the finite
lattice spacing, additional complications for the investigation of
supersymmetric theories on the lattice arise from the finite extent of
the lattice. One problem concerns for example supersymmetry breaking
due to finite temperature, or the tunneling between separate ground
states on finite volumes. While the former problem can be circumvented
by assigning periodic boundary conditions to the fermionic variables
in (imaginary-)time direction (at the price of losing the concept of
temperature), the latter problem requires an explicit extrapolation to
the thermodynamic infinite volume limit. Whether and how such an
extrapolation interferes with the extrapolation to the continuum
limit, where the lattice spacing goes to zero, is obviously an
interesting question. It is hence important to understand all the
systematics of the lattice regularisation in detail, in particular the
interplay between the infrared and ultraviolet regulators, and a
thorough comprehension of these problems and the corresponding
solutions is crucial for any investigation of spontaneous
supersymmetry breaking.

It turns out that even a simple system such as $\Ncal = 2$
supersymmetric quantum mechanics subsumes all the complications
discussed above \cite{Giedt:2004qs, Giedt:2004vb}. In addition, it
also provides a testing ground for any new approach to regularise, and
possibly simulate, supersymmetric field theories on the lattice
\cite{Wosiek:2002nm, Wosiek:2004yg, Campostrini:2002mr}. Therefore,
besides being worth studying in its own right, supersymmetric quantum
mechanics provides an ideal set up for nonperturbative investigations
of supersymmetric field theories on the lattice.  Consequently,
supersymmetric quantum mechanics on the lattice has been the subject
of intensive studies. Over time, different discretisation schemes have
been developed in order to meet the requirement of the correct
continuum limit of the theory \cite{Catterall:2000rv,
  Catterall:2009it, Catterall:2010jh}. In the context of unbroken
supersymmetry, these schemes have well established numerical support
\cite{Beccaria:1998vi, Bergner:2007pu, Kastner:2007gz,
  Bergner:2009vg}. For broken supersymmetry, however, the model
reveals a severe fermion sign problem affecting simulations with
standard Monte Carlo methods \cite{Kanamori:2007yx,
  Kanamori:2007ye}. Because of this additional obstruction, first
results in the context of broken supersymmetry were published only
very recently \cite{Wozar:2011gu}.

In a series of three papers we introduce and exploit a novel approach
with which it is possible to study, and in fact solve, supersymmetric
quantum mechanics on the lattice for both broken and unbroken
supersymmetry. In particular, we reformulate the system and its
degrees of freedom in terms of fermionic and bosonic bond
variables. This reformulation -- the subject matter of the present
paper -- is based on the exact hopping expansion of the bosonic and
fermionic actions on the lattice and allows the explicit decomposition
of the partition function into bosonic and fermionic
contributions. This explicit separation of the system paves the way
for circumventing the fermion sign problem which appears for broken
supersymmetry due to the vanishing of the Witten index. Furthermore,
the formulation in terms of bond variables enables the construction of
explicit transfer matrices which in turn allow to solve the lattice
system exactly. As a consequence we are then able to study in extenso
the continuum and infinite volume limit of systems both with broken or
unbroken supersymmetry. In particular, by means of Ward identities one
can precisely illustrate how supersymmetry is restored. Furthermore,
in the context of broken supersymmetry the emergence of the Goldstino
mode in the thermodynamic limit and at zero temperature can be studied
in detail. In summary, all the problems and issues appearing in the
context of realising supersymmetry on the lattice can be addressed and
studied by means of the exact results from the loop formulation. This
investigation will be the subject matter of the second paper in the
series.  Finally, the formulation also forms the basis for a highly
efficient fermion string algorithm \cite{Prokof'ev:2001,Wenger:2008tq}
which may be employed in numerical Monte Carlo simulations. Thus in
the third paper of the series we eventually describe the details and
properties of the algorithm which can be validated using the exact
results from the transfer matrices.  While the exact solution of the
lattice system is specific to the low dimensionality and the
subsequent simplicity of the supersymmetric quantum mechanics system,
the bond formulation and the fermion string algorithm is applicable
also to more complicated systems, e.g.~in higher dimensions, or
involving gauge fields. In particular it can be applied to
supersymmetric Yang-Mills quantum mechanics \cite{Steinhauer:2014oda}
and certain two-dimensional supersymmetric field theories, such as the
${\cal N}=1$ Wess-Zumino model
\cite{Baumgartner:2011cm,Baumgartner:2011jw,Baumgartner:2013ara,Steinhauer:2014yaa}
and the supersymmetric nonlinear O$(N)$ sigma model
\cite{Steinhauer:2013tba}.

The present paper concerns the reformulation of supersymmetric quantum
mechanics on the lattice in terms of bosonic and fermionic
bonds. Starting from the formulation of supersymmetric quantum
mechanics as an Euclidean quantum field theory, we discuss its lattice
formulation using different variants of Wilson fermions including a
$Q$-exact discretisation in section \ref{sec:SUSY_QM_lattice}. There
we also emphasise the generic fermion sign problem which arises for
numerical simulations of systems with broken supersymmetry due to the
vanishing of the Witten index. In section \ref{sec:loop formulation}
we derive the loop formulation for both the fermionic and the bosonic
degrees of freedom, while in section \ref{sec:observables} we discuss
in detail how observables such as the fermionic and bosonic two-point
functions are calculated for generic boundary conditions in the loop
formulation. Finally, in appendix \ref{app:discretisations} we
summarise the explicit actions emerging for the various
discretisations from the different superpotentials which we employ
throughout this and the following papers of the series.

\section{Supersymmetric quantum mechanics on the lattice}\label{sec:SUSY_QM_lattice}
We start our discussion with the partition function of a zero
dimensional supersymmetric quantum mechanical system with temporal
extent $L$ in the path integral formalism \cite{Creutz:1981},
\be
Z = \int {\cal D}\phi {\cal D}\psibar {\cal D}\psi \,\, \e^{-S(\phi,\psibar,\psi)} 
\ee
with the Euclidean action 
\begin{equation}\label{S_continuum}
 S(\phi,\psibar,\psi) = \int_0^\beta dt \Big\{ \frac{1}{2} \left(\frac{d\phi(t)}{dt} \right)^2 + \frac{1}{2}P^\prime(\phi(t))^2 + \psibar(t)\left( \partial_t + P^{\prime \prime}(\phi(t)) \right)\psi(t) \Big\}.
\end{equation}
Here, $\phi(t)$ is a commuting bosonic coordinate while the two
(independent) anticommuting fermionic coordinates are denoted by
$\psibar(t)$ and $\psi(t)$. The derivative of the arbitrary
superpotential $P(\phi(t))$ is taken with respect to $\phi$,
i.e.~$P^\prime = \frac{\partial P}{\partial \phi}$ and
$P^{\prime\prime} = \frac{\partial^2 P}{\partial \phi^2}$.  For
infinite temporal extent and fields vanishing at infinity, the action
is invariant under the $\Ncal = 2$ supersymmetry transformations
$\delta_{1,2}$,
\begin{equation}\label{susy_transf_cont}
\begin{array}{rclcrcl}
\delta_1 \phi &=& \overline{\epsilon} \psi, &\quad& \delta_2 \phi &=& \psibar \epsilon, \\ 
\delta_1 \psi &=& 0, &\quad&  \delta_2 \psi &=&  \left( \dot{\phi} - P^\prime \right) \epsilon,\\
\delta_1 \psibar &=& -\overline{\epsilon} \left( \dot{\phi} + P^\prime \right), &\quad&  \delta_2 \psibar &=& 0,
\end{array}
\end{equation}
where $\overline{\epsilon}$ and $\epsilon$ are Grassmann parameters
and $\dot{\phi} = \frac{d \phi}{dt}$. For finite extent, however, the
variation of the action under the supersymmetry transformations
$\delta_{1,2}$ yields the nonvanishing terms
\begin{eqnarray}\label{eq:variation_S_cont}
 \delta_1 S &  = & \int_0^\beta dt \left( -\overline{\epsilon} \left( \psi P^{\prime \prime} \dot{\phi} + \dot{\psi} P^\prime \right) \right) \, \, =  \, \, \overline{\epsilon} \psi P^\prime  \,\Big \vert_0^\beta, \\
 \delta_2 S &  = & \int_0^\beta dt \left(\dot{\psibar} \dot{\phi}  + \psibar \ddot{\phi} \right)\epsilon \, \,  =  \, \, \psibar \dot{\phi} \epsilon \, \Big\vert_0^\beta
\end{eqnarray}
which can only be brought to zero by imposing periodic boundary
conditions for the fermionic degrees of freedom, i.e.,
\begin{equation}
 \psi(\beta) = \psi(0), \qquad \psibar(\beta) = \psibar(0).
\end{equation}
Thus, choosing thermal, i.e., antiperiodic boundary conditions for the
fermionic degrees of freedom breaks supersymmetry explicitly.

For specific choices of the superpotential $P(\phi)$ the
supersymmetric system may enjoy additional symmetries.  With the
superpotential
\be
\label{eq:superpotential P_u}
P_u(\phi) = \frac{1}{2}\mu \phi^2 + \frac{1}{4}g \phi^4
\ee
the resulting action is for example invariant under a parity
transformation $\phi \rightarrow -\phi$, since
\begin{equation}
 \left( P_u^\prime(-\phi) \right)^2 = \left( P_u^\prime(\phi)\phantom{\tilde{\phi}} \!\!\! \right)^2, \qquad  P_u^{\prime \prime}(-\phi) = P_u^{\prime \prime}(\phi),
\end{equation}
and thus has an additional $\mathbbm{Z}_2$-symmetry.  This is the
potential we will use in the following as an illustrating example for
a quantum mechanical system with unbroken supersymmetry, hence the
subscript $u$. Using the superpotential
\be
\label{eq:superpotential P_b}
P_b(\phi) = -\frac{\mu^2}{4 \lambda}\phi + \frac{1}{3} \lambda
\phi^3
\ee
which we will use as an illustrating example for a system with broken
supersymmetry, one finds that the action is invariant under a combined
$CP$ symmetry,
\begin{eqnarray}\label{acc_symm_b}
 \phi(t) & \rightarrow & -\phi(t), \\
\psi(t) & \rightarrow & \psibar(t),\\
\psibar(t) & \rightarrow & \psi(t).
\end{eqnarray}
In the Schroedinger formalism, the partner potentials $\frac{1}{2}
P_b'^2 \pm P_b''$ of a system with broken supersymmetry are connected
through a mirror symmetry, and it turns out that the combined $CP$
symmetry is just a manifestation of this mirror symmetry in the field
theory language.

We now formulate the theory on a discrete lattice $\Lambda$ by
replacing the continuous (Euclidean) time variable $t \in [0,L]$ by a
finite set of $L_t$ lattice sites $x_n = a n, \, n=0,\ldots,L_t-1$
separated by the lattice spacing $a = \frac{L}{L_t}$,
\begin{equation}
 \Lambda = \{ x \in a \mathbb{Z} \ | \ 0 \leq x \leq a(L_t - 1) \}.
\end{equation}
Then, in order to formulate the path integral of supersymmetric
quantum mechanics as a one-dimensional lattice field theory, we define
the path integral measure on the lattice as
\begin{equation}
 \int \mathcal{D}\phi \mathcal{D}\psibar \mathcal{D}\psi 
\equiv
\prod_{x = 0}^{L_t - 1} \int_{-\infty}^{\infty} d\phi_x \int d\psibar_x \int d\psi_x,
\end{equation}
such that the lattice partition function is given by
\begin{equation}\label{eq:partition_function}
 Z = \int \mathcal{D} \phi \mathcal{D} \psibar\mathcal{D} \psi \
 \mathrm{e}^{-S_\Lambda(\phi,\psibar,\psi)} \, ,
\end{equation}
where $S_\Lambda$ is a suitable discretisation of the action. It
requires the replacement of the temporal integration in the action by
a discrete sum over all lattice sites,
\begin{equation}
 \int_0^L dt \, \longrightarrow \, a\sum_{x = 0}^{L_t - 1},
\end{equation}
and the replacement of the continuous derivatives by suitable lattice
derivatives. In the following two subsections
\ref{subsec:latt_stand_dis} and \ref{subsec:Q-exact discretisation} we
discuss in detail two suitable lattice actions.

In principle, it is now straightforward to evaluate the partition
function (\ref{eq:partition_function}), for example numerically using
Monte Carlo algorithms. However, for a system with broken
supersymmetry one encounters a severe fermion sign problem when
standard Metropolis update algorithms are employed. We will address
this issue in more detail in subsection
\ref{subsec:fermion_sign_problem}.

Finally, we note that the continuum limit of the lattice theory is
taken by fixing the dimensionful parameters $\mu, g, \lambda$ and $L$
while taking the lattice spacing $a \rightarrow 0$. In practice, the
dimensionless ratios $f_u = g/\mu^2, f_b = \lambda/\mu^{3/2}$ fix the
couplings and $\mu L$ the extent of the system in units of $\mu$,
while $a \mu$ and $a/L$ are subsequently sent to zero. Then, by
attaching a physical scale to $L$ for example, the physical values for
all other dimensionful quantities follow immediately. Employing
antiperiodic boundary conditions for the fermion, the extent $L$
corresponds to the inverse temperature, hence the system at finite
$\mu L$ represents a system at finite temperature and the limit $\mu L
\rightarrow \infty$ implies a system at zero temperature.

\subsection{Standard discretisation}\label{subsec:latt_stand_dis}
The most obvious choice for discretising the continuous derivatives in
the action is to use the discrete symmetric derivative
\be 
\widetilde\nabla  =  \frac{1}{2}(\nabla^+ + \nabla^-)
\ee
where
\begin{eqnarray}\label{lat_operators_1d}
 \nabla^-f_x & = & \frac{1}{a}(f_x - f_{x-a}), \\
 \nabla^+f_x & = & \frac{1}{a}(f_{x+a} - f_x)
\end{eqnarray}
are the backward and forward derivatives, respectively. However, it is
well known that the symmetric derivative leads to the infamous fermion
doubling which, for the sake of maintaining supersymmetry, should be
avoided. This can be achieved by introducing an additional Wilson term
which removes all fermion doublers from the system,
\begin{displaymath}
 \nabla^W(r) = \widetilde \nabla - \frac{r a}{2} \Delta,
\end{displaymath}
where $\Delta = \nabla^+ \nabla^-$ is the Laplace operator and the
Wilson parameter takes values \mbox{$r \in [-1,1] \backslash \{0 \}$}. It
turns out that for one-dimensional derivatives the standard choice $r
= \pm 1$ yields $\nabla^W(\pm 1) = \nabla^{\mp}$, hence for $r=1$ the
discretised action reads
\begin{equation}\label{eq:S_stand_a}
 S_\Lambda = a\sum_x  \Big\{ \frac{1}{2}(\nabla^-\phi_x)^2 + \frac{1}{2}P^\prime(\phi_x)^2 + \psibar_x(\nabla^- + P^{\prime \prime}(\phi_x))\psi_x \Big\}
\end{equation}
and setting the lattice spacing $a = 1$ we obtain
\begin{equation}\label{eq:S_stand}
  S_\Lambda = \sum_x \Big\{ \frac{1}{2}(P^\prime(\phi_x)^2 + 2\phi_x^2) - \phi_x\phi_{x-1} + (1 + P^{\prime \prime}(\phi_x))\psibar_x\psi_x - \psibar_x\psi_{x-1 } \Big\}.
\end{equation}
This is the standard discretisation for the action of supersymmetric
quantum mechanics on the lattice.  Correspondingly, the supersymmetry
transformations (\ref{susy_transf_cont}) discretised on the lattice
$\Lambda$ read
\begin{equation}\label{susy_transf_latt}
\begin{array}{rclcrcl}
\delta_1 \phi &=& \overline{\epsilon} \psi, &\quad& \delta_2 \phi &=& \psibar \epsilon, \\ 
\delta_1 \psi &=& 0, &\quad&  \delta_2 \psi &=&  \left( \nabla^- \phi - P^\prime \right) \epsilon,\\
\delta_1 \psibar &=& -\overline{\epsilon} \left( \nabla^- \phi  + P^\prime \right), &\quad&  \delta_2 \psibar &=& 0,
\end{array}
\end{equation}
and the variation of the action under $\delta_1$ yields
\be \label{eq:varitation_1_S_lattice}
 \delta_1 S_\Lambda = -\overline{\epsilon} \sum_x \left\{ \psi_x
   P^{\prime \prime}(\phi_x) (\nabla^- \phi_x) + P^\prime(\phi_x)
   (\nabla^- \psi_x) \right\} \, ,
\ee
and similarly for $\delta_2$. Note, that
(\ref{eq:varitation_1_S_lattice}) is the lattice version of the
surface term in the continuum, eq.(\ref{eq:variation_S_cont}). Since
the Leibniz rule does not apply on the lattice, it is not possible to
integrate this term by parts and $S_\Lambda$ is therefore not
invariant under the supersymmetry transformations $\delta_1$ and
$\delta_2$. This is the explicit supersymmetry breaking by the lattice
discretisation which we already pointed out in the introduction.  In
addition, the Wilson term also breaks the time reversal symmetry, or
equivalently the charge conjugation for the fermion in our quantum
mechanical system. This can be seen from the fact that the oriented
hopping term $\psibar_x \psi_{x-1}$ is directed only in forward
direction $x-1 \rightarrow x$, while the backward hopping is
completely suppressed\footnote{For an arbitrary choice of the Wilson
  parameter $0 < |r| < 1$ both directions would be present.}. As a
matter of fact, the discretised system only describes a fermion
propagating forward in time, but not the corresponding antifermion
propagating backward in time. As we will see later, this has an
important consequence for the fermion bond formulation.  In the
continuum the symmetry is restored and this comes about by the
relative contributions of the fermion and antifermion approaching each
other in this limit.

\begin{figure}[tbh]
  \centering
 \includegraphics[width = 0.8\textwidth]{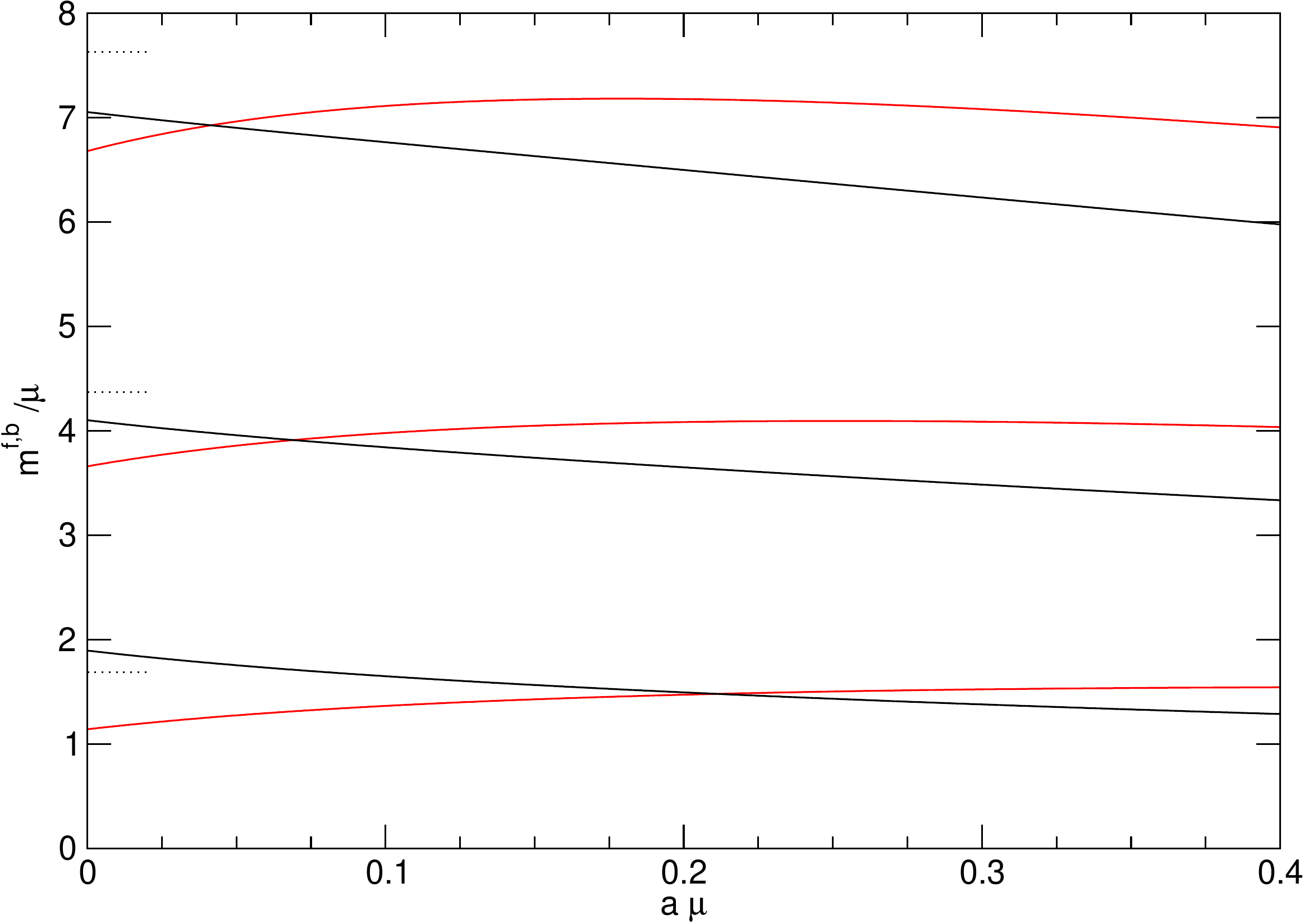}
 \caption{ Continuum extrapolation $a\mu \rightarrow 0$ of the bosonic
   (red lines) and fermionic (black lines) mass gaps expressed in
   units of $\mu$ for unbroken supersymmetry at the coupling $f_u = 1$
   using the standard discretisation. The expected continuum values
   are indicated by dotted lines.}
  \label{fig:continuum_masses_stand}
\end{figure}

At this point, it is necessary to stress that the action in
eq.(\ref{eq:S_stand_a}) does not correctly reproduce the continuum
limit of the theory
\cite{Catterall:2000rv,Giedt:2004vb,Bergner:2007pu}.  In figure
\ref{fig:continuum_masses_stand}, we illustrate this failure by
extrapolating the lowest mass gaps of the fermion and the boson for
the system with superpotential $P_u$ (unbroken supersymmetry) to the
continuum $a\mu \rightarrow 0$. The exact calculation is based on the
extraction of the mass gaps via transfer matrix techniques which will
be discussed in detail in the second paper of this series
\cite{Baumgartner:2015qba}, see also \cite{Baumgartner:2012np}. Note,
that the extrapolation of the masses does not yield the known
continuum values indicated by the horizontal dotted lines. In fact the
bosonic and fermionic mass gaps are not even degenerate in the
continuum and supersymmetry is not restored for this discretisation.
It turns out that the mismatch is due to perturbative corrections and
a careful analysis of those in the lattice theory is therefore
mandatory \cite{Giedt:2004vb}. However, since this quantum mechanics
model is superrenormalisable, there are only a finite number of terms
which do not converge to the correct continuum limit, and it is
therefore sufficient to add a finite number of counterterms to the
lattice action. Note that as opposed to a quantum field theory in
higher dimensions, the counterterms do not diverge in quantum
mechanics, but remain finite as $a\rightarrow 0$. As explicitly shown
in \cite{Giedt:2004vb}, in order to restore supersymmetry in the
continuum, it is necessary and sufficient to add the term $ P^{\prime
  \prime}/2$ to the lattice action, i.e.,
\be
\label{eq:action_counterterm}
S_\Lambda \, \longrightarrow \,  S_\Lambda^c =  S_\Lambda + \frac{1}{2} \sum_x P^{\prime
  \prime}(\phi_x) \, .
\ee
The term can be understood as a radiative correction and we will see
in section \ref{subsec:fermion_sign_problem} how the term arises in
the explicit calculation of the determinant of the Wilson Dirac
matrix.  Finally, it is important to note that the resulting lattice
theory is not supersymmetric at finite lattice spacing, but in the
continuum limit it will nevertheless flow to the correct
supersymmetric theory without any further fine tuning.

\subsection{$Q$-exact discretisation}
\label{subsec:Q-exact discretisation}
A discretisation of supersymmetric quantum mechanics which avoids the
fine tuning of counter\-terms is based on the idea that it might be
sufficient to preserve only a subset of the full supersymmetry at
finite lattice spacing in order to reach the correct continuum
limit. This approach, known under the name of twisted supersymmetry,
was first applied to supersymmetric quantum mechanics in
\cite{Catterall:2000rv} and can be established in the context of
topological field theory \cite{Catterall:2003wd}, or from a lattice
superfield formalism \cite{Giedt:2004qs}. For ${\cal N}=2$
supersymmetric quantum mechanics it relies on the observation that the
lattice variation $\delta_1 S_\Lambda$ of the standard discretised
action in (\ref{eq:varitation_1_S_lattice}) can be written -- up to a
minus sign -- as the variation of the lattice operator
\begin{equation}
\label{eq:Qexact_term}
 O = \sum_x P^\prime(\phi_x)(\nabla^- \phi_x)
\end{equation}
under the same supersymmetry transformation $\delta_1$, such that we have
\begin{equation}
 \delta_1 S_\Lambda = - \delta_1 O.
\end{equation}
It is then clear that the invariance of the action under the
supersymmetry transformation $\delta_1$ can be restored by simply
adding the term $O$ to the action.  The bosonic part of the so
constructed action can be written as
\begin{equation}\label{eq:action_Q1}
 S_{\Lambda,\text{bosonic}}^{Q} = \sum_x \Big\{ \frac{1}{2}(\nabla^-\phi_x)^2 +
 \frac{1}{2}P^\prime(\phi_x)^2 
+ P^\prime(\phi_x)(\nabla^- \phi_x) \Big\} \, ,
\end{equation}
and the total action in more  compact form as
\begin{equation}\label{eq:action_Q1_compact}
  S_{\Lambda}^{Q} = \sum_x \Big\{\frac{1}{2} \left( (\nabla^-\phi_x) +
    P^\prime(\phi_x) \right)^2 
+ \psibar_x(\nabla^- + P^{\prime \prime}(\phi_x))\psi_x \Big\}\, .
\end{equation}
This is the $Q$-exact lattice action which preserves the supersymmetry
$\delta_1$ exactly (but not $\delta_2$) for finite lattice
spacing. The $Q$-exactness can be best seen in the off-shell
formulation of the total action. Using an auxiliary field and defining
the fermionic variation by $\delta_1 = \overline{\epsilon}Q$, where
$Q$ is the generator of the supersymmetry transformation
\cite{Catterall:2003wd}, one can write the total action off-shell as
the $Q$-variation of a particular function $F$, i.e.~$S_\Lambda = Q
F$. This makes the $Q$-exact invariance of the action explicit via the
nilpotency property $Q^2=0$ \cite{Catterall:2009it}.
Maintaining this single supersymmetry on the lattice is sufficient to
protect the theory from radiative contributions which would otherwise
spoil the continuum limit. Note, that this action corresponds to the
Ito prescription in \cite{Bergner:2007pu}.  In complete analogy, one
can also construct a $Q$-exact action invariant under $\delta_2$ but
not $\delta_1$, or in fact a $Q$-exact action invariant under any
linear combination of $\delta_1$ and $\delta_2$, but not invariant
under the corresponding orthogonal linear combination. This property
is related to the fact that the improved lattice field theory is
topological and hence the improvement term $O$ can be added to the
action with any prefactor different from zero to obtain a $Q$-exact
action \cite{Catterall:2003wd}. Each variant leads to a different
discretisation of the bosonic part of the action. For the loop
formulation we will concentrate on the form given in
eq.(\ref{eq:action_Q1}) and (\ref{eq:action_Q1_compact}), but of
course the reformulation can be achieved for any $Q$-exact
action. Before getting more specific, we will now discuss the fermion
sign problem emerging in simulations of systems with broken
supersymmetry.

\subsection{Fermion sign problem from broken supersymmetry}
\label{subsec:fermion_sign_problem}
In this section we discuss the fermion sign problem which affects
standard Monte Carlo simulations of supersymmetric systems with broken
supersymmetry. The problem is generic and affects all supersymmetric
systems with (spontaneously) broken supersymmetry since it is related
to the vanishing of the Witten index accompanying any spontaneous
supersymmetry breaking. In the particularly simple supersymmetric
quantum mechanics case we consider here in this paper, the problem can
be illustrated very explicitly.

In order to evaluate the partition function in
eq.(\ref{eq:partition_function}), in a first step one usually
integrates out the fermionic degrees of freedom in the path integral
which then yields the determinant of the fermion Dirac operator
$D(\phi)$, i.e.,
\begin{equation}\label{eq:exp_val_det_D}
 Z = \int \mathcal{D}\phi \ \det (D(\phi)) \ \e^{-S^B_\Lambda(\phi)},
\end{equation}
with $S_\Lambda^B(\phi)$ being the purely bosonic part of the lattice
action. In the following we will concentrate on the Wilson Dirac
operator $D(\phi) = \nabla^- + P^{\prime\prime}(\phi)$, but the
considerations apply equally to any fermion discretisation.  It turns
out that depending on the specific choice of the superpotential
$P(\phi)$ the determinant is not positive definite. In that case the
effective Boltzmann weight $\det (D(\phi)) \exp\{-S_\Lambda^B(\phi)\}$
cannot be interpreted as a probability distribution and the standard
Monte Carlo approach breaks down. In fact, since the partition
function with periodic boundary conditions is proportional to the
Witten index, which vanishes in systems with (spontaneously) broken
supersymmetry, the partition function itself must be zero. From
eq.(\ref{eq:exp_val_det_D}) it then becomes clear that this can only
be achieved by the determinant being indefinite and in fact zero on
average. The cancellations between positive and negative contributions
of the determinant to the partition function are hence maximal and
constitute a severe fermion sign problem.  Since the fermion
determinant $\det(D(\phi))$ can be calculated analytically both in the
continuum \cite{Cooper:1994eh,Bergner:2007pu,PhysRevD.28.1922} and on
the lattice, one can illustrate this explicitly and we will do so in
the next two subsections. Moreover, the considerations will also be
useful for the interpretation of the reformulation in terms of fermion
loops.

\subsubsection{The fermion determinant in the continuum}
For the evaluation of the fermion determinant in the continuum, some
regularisation is necessary. A suitable choice is given by dividing
the determinant by the fermion determinant of the free theory,
$\det(\partial_t + \mu)$. Moreover, the computation of the fermion
determinant depends essentially on the choice of the boundary
conditions for the fermionic degrees of freedom.

For antiperiodic boundary conditions $\psi(L) = -\psi(0)$, the
regularised determinant yields
\begin{equation}\label{eq:det_ferm_apbc}
 \det \left( D(\phi) \right) \doteq \det \left( \frac{\partial_t +
     P^{\prime \prime}(\phi) }{\partial_t + \mu }\right) = \frac{\cosh
   \left(\frac{1}{2}\int_0^L dt\  P^{\prime \prime}(\phi)
   \right)}{\cosh \left( \frac{1}{2} \mu L \right)}
\end{equation}
and we observe that this is always positive. Furthermore, writing
the cosh function in terms of exponentials, we find that
\begin{equation}\label{fermion_det}
 \det \left( D(\phi) \right) \propto \exp \left( +\frac{1}{2} \int_0^L dt \ P^{\prime \prime}(\phi) \right) + \exp \left( -\frac{1}{2} \int_0^L dt \ P^{\prime \prime}(\phi) \right)
\end{equation}
and hence the partition function (\ref{eq:exp_val_det_D})
decomposes into two parts which just correspond to the bosonic and the
fermionic sector, respectively. To be specific, one has
\begin{equation}
 \int \D \phi \ \det \left( D(\phi) \right) \ \e^{-S^B(\phi)} = \int 
 \D \phi \ \e^{-S_-^B(\phi)} + \int 
 \D \phi \ \e^{-S_+^B(\phi)}\equiv Z_0 + Z_1 \, ,
\end{equation}
where the actions
\begin{equation}
S_\pm^B(\phi) = \int_0^L dt \Big\{ \frac{1}{2}
  \left(\frac{d\phi(t)}{dt} \right)^2 + \frac{1}{2}P^\prime(\phi(t))^2
  \pm \frac{1}{2} P^{\prime \prime}(\phi(t)) \Big\}
\end{equation}
remind us of the partner potentials in the usual Hamilton formulation
of supersymmetric quantum mechanics, and $Z_0$ and $Z_1$ are the
partition functions in the bosonic and fermionic sector,
respectively. Since we have calculated the determinant for
antiperiodic boundary conditions, we have
\begin{equation}
Z_0 + Z_1 = Z_{a}
\end{equation}
and we note that $Z_{a}$ is positive since both $Z_0$ and $Z_1$ are
positive.

For periodic boundary conditions $\psi(L) = \psi(0)$, the
analogous calculation of the regularised fermion determinant yields
\begin{equation}\label{eq:det_ferm_pbc}
 \det \left( D(\phi) \right) = \frac{\sinh \left(\frac{1}{2}\int_0^L
     dt \ P^{\prime \prime}(\phi) \right)}{\sinh \left( \frac{1}{2}
     \mu L \right)}
\end{equation}
and writing out the sinh function as a sum of exponentials, we find
\begin{equation}\label{eq:det_Z_pbc}
\int \D \phi \ \det \left( D(\phi) \right) \ \e^{-S^B(\phi)} = \int 
 \D \phi \ \e^{-S_-^B(\phi)} - \int 
 \D \phi \ \e^{-S_+^B(\phi)} = Z_0 - Z_1 \equiv Z_{p} \,.
\end{equation}
More importantly, we note that for this choice of boundary conditions
the partition function is indefinite and the fermion determinant is
hence not necessarily positive.

We now recalling the definition of the Witten index $W$ from quantum
mechanics \cite{Witten:1982},
\begin{equation}
\label{eq:QM_Witten_index}
W =  \Tr \left[(-1)^F e^{- \beta H} \right]= \Tr_B  \left[e^{-\beta H}\right] - \Tr_F  \left[e^{-\beta H}\right]
\end{equation}
where $H$ is the Hamilton operator and $F$ the fermion number, while
$\Tr_{B,F}$ denotes the trace over the bosonic and fermionic states,
respectively. Identifying $\beta$ with $L$ we realise that the Witten
index is in fact proportional to the expectation value of the fermion
determinant, i.e., the partition function with fully periodic boundary
conditions,
\begin{equation}\label{WI_prop_Zp}
  W \propto Z_{p}\,.
\end{equation}
The relation is given as a proportionality because the path integral
measure is only defined up to a constant multiplicative factor as
compared to the traces in eq.(\ref{eq:QM_Witten_index}).

In order to see the implications of these results, we consider the two
superpotentials $P_u$ and $P_b$ defined in the introduction of section
\ref{sec:SUSY_QM_lattice}. Recall that the superpotential $P_u$ in
eq.(\ref{eq:superpotential P_u}) is invariant under the parity
transformation $\phi \rightarrow - \phi$. Furthermore, for $\mu > 0$
and $g \geq 0$, $P_u^{\prime \prime}(\phi) > 0$, and
eq.(\ref{eq:det_ferm_pbc}) and (\ref{eq:det_Z_pbc}) then imply that
$Z_{p} \neq 0$ and hence the Witten index is nonzero, $W \neq
0$. Thus, we conclude that for this superpotential supersymmetry is
indeed unbroken, in agreement with the generic expectation from
supersymmetric quantum mechanics. Next, we consider the superpotential
$P_b$ in eq.(\ref{eq:superpotential P_b}) which we recall is odd under
the parity transformation $\phi \rightarrow - \phi$, and so is its
second derivative, $P_b^{\prime \prime}(- \phi) \rightarrow -
P_b^{\prime \prime}(\phi)$. On the other hand, the bosonic action
$S^B(\phi)$ for this superpotential,
\begin{equation}
 S^B(\phi) = \int dt \ \Big\{ \frac{1}{2}\left( \frac{d \phi}{dt} \right)^2
 - \frac{1}{2} \left( \frac{\mu^2}{2}\phi^2 - \lambda^2 \phi^4 \right) \Big\}
 \, ,
\end{equation}
is invariant under the parity transformation, $S^B(-\phi) \rightarrow
S^B(\phi)$. Therefore, eq.(\ref{eq:det_ferm_pbc}) and
(\ref{eq:det_Z_pbc}) imply that with periodic b.c.~for each
configuration contributing to the partition function, there is the
parity transformed one with exactly the same weight but opposite sign
coming from the fermion determinant. Consequently, the partition
function $Z_{p}$ vanishes and the Witten index is $W = 0$. Indeed, for
the superpotential $P_b$ one expects on general grounds that
supersymmetry is broken.

Obviously, the argument can be reversed leading to the conclusion
discussed at the beginning of this section: since the Witten index is
zero for a supersymmetric system with broken supersymmetry, the
partition function with periodic boundary conditions $Z_{p}$, and
hence the expectation value of $\det(D)$, vanishes, and this then
leads to the fermion sign problem for numerical simulations.

\subsubsection{The fermion determinant on the lattice}
\label{subsubsec:lattice determinant}
Next, we calculate the fermion determinant on the lattice. The lattice
provides a regularisation, such that we can calculate the determinant
directly without division by the determinant of the free theory. Using
the lattice discretisation introduced in section
\ref{subsec:latt_stand_dis}, the determinant of the fermion matrix can
easily be seen to be
\begin{equation}\label{eq:ferm_det_lat}
 \det \left( \nabla^- + P^{\prime \prime}(\phi_x) \right) =  \prod_x (1 + P^{\prime \prime}(\phi_x) ) \mp 1,
\end{equation}
where the $-1$ $(+1)$ in the last term is associated with periodic
(antiperiodic) boundary conditions. Note that this result is
consistent with the expression derived for supersymmetric Yang-Mills
quantum mechanics in \cite{Steinhauer:2014oda}. As in the continuum
the fermion determinant decomposes into a bosonic part, the product
over all lattice sites $x$, and a fermionic part, the term $\mp 1$. We
will see later in section \ref{sec:loop formulation} from the fermion
loop formulation that this interpretation is indeed correct.

At this point it is interesting to discuss the continuum limit of the
lattice determinant. In principle, one would expect to recover the
expressions in eq.(\ref{eq:det_ferm_apbc}) and
eq.(\ref{eq:det_ferm_pbc}) when dividing the lattice determinant by
the determinant of the free lattice theory and then taking the lattice
spacing to zero, $a \rightarrow 0$. However, one finds
\begin{equation}\label{lattice_det_a_to_0}
 \lim_{a \rightarrow 0} \det \left( \frac{\nabla^- + P^{\prime
       \prime}(\phi_x)}{\nabla^- + \mu \cdot \mathds{1}} \right) =
 \frac{\exp \left( \frac{1}{2} \int_0^L dt \ P^{\prime
       \prime}(\phi) \right)}{\exp \left( \frac{1}{2} \mu L
   \right)} \det \left( \frac{\partial_t + P^{\prime
       \prime}(\phi)}{\partial_t + \mu} \right) \, ,
\end{equation}
i.e., taking the naive continuum limit apparently yields an additional
factor in front of the continuum determinant. This factor can be
understood as the remnants of the radiative corrections from the
Wilson discretisation which survive the naive continuum limit
\cite{Giedt:2004vb}. The term is in fact responsible for the wrong
continuum limit of the fermion and boson masses discussed in section
\ref{subsec:latt_stand_dis} and illustrated in figure
\ref{fig:continuum_masses_stand}.

Let us now proceed by discussing the determinant of the Wilson Dirac
matrix for both superpotentials $P_u$ and $P_b$ explicitly.  Using the
superpotential for unbroken supersymmetry $P_u$, the determinant
yields
\begin{equation}
 \det(\nabla^- + P^{\prime \prime}_u(\phi_x)) = \prod_x(1 + \mu +
 3g\phi^2_x) \mp 1 
\end{equation}
which for $\mu > 0$ and $g \geq 0$ is strictly positive, independent
of the boundary conditions.  Using the superpotential for broken
supersymmetry $P_b$, the determinant yields
\begin{equation}\label{eq:lattice determinant Pb}
 \det(\nabla^- + P^{\prime \prime}_b(\phi_x)) = \prod_x(1 + 2 \lambda \phi_x) \mp 1
\end{equation}
which is indefinite even for $\lambda > 0$. While this is necessary in
order to accommodate a vanishing Witten index, it imposes a serious
problem on any Monte Carlo simulation, for which positive weights, and
hence positive determinants, are strictly required. Moreover, the sign
problem is severe in the sense that towards the continuum limit (i.e.,
when the lattice volume goes to infinity), the fluctuations of the
first summand in eq.(\ref{eq:lattice determinant Pb}) around 1 tend to
zero, such that $W \rightarrow 0$ is exactly realised in that
limit. Hence, the source of the fermion sign problem lies in the exact
cancellation between the first and the second summand in
eq.(\ref{eq:lattice determinant Pb}), i.e., of the bosonic and
fermionic contributions to the partition function, and this
observation also holds more generally in higher dimensions
\cite{Baumgartner:2011cm,Baumgartner:2011jw,Baumgartner:2013ara}.  In
the loop formulation, to be discussed in the next section, the
separation of the partition function into the various fermionic and
bosonic sectors is made explicit and allows the construction of a
simulation algorithm that samples these sectors separately, and more
importantly also samples the relative weights between them. In this
way, the loop formulation eventually provides a solution to the
fermion sign problem.

\section{Loop formulation of supersymmetric quantum mechanics}
\label{sec:loop formulation}
We will now discuss in detail the reformulation of supersymmetric
quantum mechanics in terms of bosonic and fermionic bonds, eventually
leading to the so-called loop formulation. The bond formulation is
based on the hopping expansion for the bosonic and fermionic degrees
of freedom. For the latter, the hopping expansion becomes particularly
simple due to the nilpotent character of the fermionic variables and
in addition reveals the decomposition of the configuration space into
the bosonic and fermionic subspaces.

\subsection{Loop formulation of the fermionic degrees of freedom}
\label{subsec:fermion loop formulation}
\begin{figure}[h]
\centering
\includegraphics[width=0.16\textwidth]{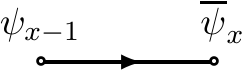}
 \caption{Graphical representation of the directed fermionic bond $b^F$.}
  \label{fig:fermionicbond}
\end{figure}

We start by splitting the action into a bosonic and fermionic part
\begin{align}
  S_\Lambda &= S_\Lambda^B(\phi) + S_\Lambda^F(\phi,\psibar,\psi)\\
\intertext{with}
S_\Lambda^B(\phi) & =  \sum_x \Big\{ \frac{1}{2}(\nabla^-\phi_x)^2 +
  \frac{1}{2}P^\prime(\phi_x)^2 \Big\}, \\
 S_\Lambda^F(\phi,\psibar,\psi) & =  \sum_x \Big\{\psibar_x(\nabla^- +
 P^{\prime \prime}(\phi_x))\psi_x \Big\}\, ,
\end{align}
so that the partition function can be written as
\begin{equation}\label{part_fun_path_integral}
 Z = \int \mathcal{D}\phi \ \e^{-S_\Lambda^B(\phi)}\int
 \mathcal{D}\psibar \mathcal{D}\psi \
 \e^{-S_\Lambda^F(\phi,\psibar,\psi)} \, .
\end{equation}
Rewriting the fermionic action as in eq.(\ref{eq:S_stand}) and
introducing $M(\phi) = 1 + P^{\prime \prime}(\phi)$ for the monomer
term we have
\be
  S_\Lambda^F(\phi,\psibar,\psi) = \sum_x \Big\{ M(\phi_x)\psibar_x\psi_x - \psibar_x\psi_{x-1 } \Big\}
  \, ,
\ee
and expanding separately the two terms in the Boltzmann
factor yields
\begin{equation}
 \e^{-S_\Lambda^F} = \prod_x (1 - M(\phi_x))\psibar_x \psi_x \prod_x \left(1 +
   \psibar_x \psi_{x - 1} \right) \,.
\end{equation}
Due to the nilpotency of the Grassmann variables, all terms of second
or higher order in $\psibar_x \psi_x$ or $\psibar_x \psi_{x - 1}$
vanish in the expansion.  Introducing fermionic monomer occupation
numbers $m(x) \in \{ 0,1 \}$ as well as the fermionic bond occupation
numbers $n^f(x) \in \{ 0,1 \}$, we can further rewrite the expansion
as
\begin{equation}\label{expansion_fermionic}
 \e^{-S_\Lambda^F} = \prod_x \left( \sum_{m(x) = 0}^1\left(-M(\phi_x) \psibar_x \psi_x \right)^{m(x)} \right) \prod_x \left( \sum_{n^f(x) = 0}^1 \left( \psibar_x \psi_{x - 1}\right)^{n^f(x)} \right).
\end{equation}
The fact that the fermionic occupation numbers can only take the
values 0 or 1 can be seen as a realisation of the Pauli exclusion
principle and follows naturally from the nilpotency property of the
fermion fields. Obviously, it is natural to assign the bond occupation
number $n^f(x)$ to the link connecting the sites $x-1$ and $x$, while
the monomer occupation number $m(x)$ lives on the lattice site. The
directed fermionic bond can be represented as illustrated in figure
\ref{fig:fermionicbond} by an arrow associated to the hopping term
$\psibar_x\psi_{x-1}$ which is either occupied or not.

In a next step we can now integrate out the fermionic variables. The
Grassmann integration rule
\be
\label{eq:Grassmann integration rule}
\int d\psibar d\psi \,\, \psibar \psi = - 1
\ee
tells us that each site $x$ must be occupied by exactly one variable
$\psi_{x}$ and one variable $\psibar_{x}$ in order to obtain a nonzero
contribution to the path integral. The Grassmann integration at a
given site $x$ is either saturated by the monomer term \mbox{$\propto
  \psibar_x \psi_x$}, yielding the contribution $M(\phi_x)$ after the
integration, or by exactly one ingoing and one outgoing fermionic bond
\mbox{ $\propto \psibar_{x+1} \psi_x \cdot \psibar_x\psi_{x-1}$},
yielding the contribution 1 for each bond after the Grassmann
integration. The fact that these two possibilities are exclusive at
each site leads to a local constraint on the monomer and bond
occupation numbers $m(x)$ and $n^f(x)$ given by
\begin{equation}\label{eq:ferm_cond_QM}
 m(x) + \frac{1}{2}\left( n^f(x) + n^f(x - 1) \right) = 1, \qquad
 \forall x \, .
\end{equation}
As a consequence, the integration over the Grassmann degrees of
freedom $\psi$ and $\psibar$ is replaced by a sum over all possible
configurations of monomer and bond occupation numbers satisfying the
local constraint (\ref{eq:ferm_cond_QM}).  The constraint implies
that there are only two possible fermionic bond configurations with
nonzero weight. On the one hand, eq.(\ref{eq:ferm_cond_QM}) is
satisfied if $m(x) = 1$ and $ n^f(x) = 0 \ \forall x$. In this case,
there are no fermionic bonds, i.e.~the fermion number is $F=0$, and
such a configuration hence contributes to the bosonic sector.  Each
site is then saturated with the monomer term and by applying the
Grassmann integration rules we identify the total fermionic
contribution to the weight of a configuration to be the product of
monomer weights $M(\phi_x)$ at each site $x$, i.e., $\prod_x
(1+P^{\prime \prime}(\phi_x))$.  On the other hand,
eq.(\ref{eq:ferm_cond_QM}) can also be satisfied by $n^f(x) = 1$ and
$m(x) = 0 \ \forall x$. For such a configuration the fermion number is
$F=1$, since all sites $x$ are connected by fermionic bonds forming a
fermionic loop which winds around the lattice. The fermionic bonds
contribute with weight 1, hence the total fermionic contribution to
the weight of such a configuration is just a factor $(-1)$ where the
minus sign follows from integrating out the cyclic loop of hopping
terms and is the usual, characteristic fermion sign associated with
closed fermion loops. In addition, the fermion loop receives an
additional minus sign if antiperiodic boundary conditions for the
fermion field are employed. We will discuss this in more detail in
section \ref{sec:boundary conditions}.

Summarising the two contributions to the path integral from the
integration of the fermionic variables, we have
\be
\prod_x \Big(1+P^{\prime \prime}(\phi_x)\Big) \mp 1
\ee
for periodic and antiperiodic b.c., respectively, and we recognise
this as the determinant of the lattice Dirac operator,
cf.~eq.(\ref{eq:ferm_det_lat}). The first term from the configuration
without any fermionic bonds is identified as the bosonic contribution
to the path integral, while the second term from the fermion loop
configuration is identified as the fermionic contribution. The
partition function can hence be written as
\be
Z_{p,a} = Z_0 \mp Z_1
\ee
with 
\begin{eqnarray}
Z_0 &=&  \int \D \phi \,\, \e^{-S_\Lambda^B(\phi)} \prod_x \Big(1+P^{\prime  \prime}(\phi_x)\Big) ,\\
Z_1 &=& \int \D \phi \, \,\e^{-S_\Lambda^B(\phi)}
\end{eqnarray}
where the subscript $0$ and $1$ denotes the fermion winding number of
the underlying fermionic bond configuration, or equivalently the
fermion number $F$. We have thus confirmed the interpretation of the
bosonic and fermionic parts contributing to the fermion determinant
alluded to in section \ref{subsubsec:lattice determinant}.

\subsection{Loop formulation of the bosonic degrees of freedom}
In complete analogy to the previous section we can also replace the
continuous bosonic variables $\phi$ by integer bosonic bond occupation
numbers. To keep the discussion simple we first consider the standard
discretisation.  The bosonic action $S_\Lambda^B$ in
eq.(\ref{eq:S_stand}) can be written in the form
\be
S_\Lambda^B = \sum_x \left\{ - w \cdot \phi_x\phi_{x-1} + V(\phi_x) \right\}
\ee
where we have separated the (nonoriented) hopping term $w \cdot
\phi_x\phi_{x-1}$ with the hopping weight $w=1$ from the local
potential term $V(\phi_x) = \frac{1}{2}(P^\prime(\phi_x)^2 +
2\phi_x^2)$.  Expanding now the exponential of the hopping term in the
Boltzmann factor we obtain
\begin{equation}\label{eq:expansion_b}
 \e^{-S_\Lambda^B} = \prod_x \left( \sum_{n^b(x) = 0}^\infty \frac{(w
     \cdot \phi_{x-1}\phi_x)^{n^b(x)}}{n^b(x)!} \right) \prod_x \mathrm{e}^{-V(\phi_x)}.
\end{equation}
The summation indices $n^b(x)$ can be interpreted as bosonic bond
occupation numbers, but in contrast to the fermionic case there is no
Pauli exclusion principle which truncates the expansion, and hence the
summation runs from 0 to infinity.

\begin{figure}[h]
\centering
\includegraphics[width=0.16\textwidth]{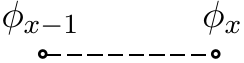}
\caption{Graphical representation of the bosonic bond $b^B_{1 \rightarrow 1}$.}
  \label{fig:bosonicbond}
\end{figure}

To make further progress we now need to combine this with the result
from the expansion in the fermionic variables, and so we obtain for
the full partition function
\begin{equation}
 Z = \int \mathcal{D} \phi \prod_x \left( \sum_{n^b(x) = 0}^\infty
   \frac{(w\cdot \phi_{x - 1}\phi_x)^{n^b(x)}}{n^b(x)!} \right) \prod_x \mathrm{e}^{-V(\phi_x)} \prod_x \left( \sum_{m(x) = 0}^1 \! \! \! \! M(\phi_x)^{m(x)} \right).
\end{equation}
In order to integrate over the variable $\phi_x$ locally at each site
we select a particular entry in each of the sums. This is equivalent
to choosing a particular bond configuration $\{n^b(x)\}$ and fermionic
monomer configuration $\{m(x)\}$. The rearrangement of the bosonic
fields, essentially collecting locally all powers of $\phi_x$, yields
local integrals of the form
\begin{equation}
 Q(N(x),m(x)) = \int_{-\infty}^\infty d \phi_x \ \phi_x^{N(x)} \e^{-V(\phi_x)}M(\phi_x)^{m(x)}
\end{equation}
where the site occupation number 
\be
N(x) = n^b(x) + n^b(x - 1)
\ee
counts the total number of bosonic bonds attached to the site
$x$. This can be visualised by a graphical representation of the bond
as a (dashed) line connecting the sites $x-1$ and $x$ as in figure
\ref{fig:bosonicbond}. The site occupation number is then just the
number of bonds connected to a site from the left and the right.

As a consequence of the reordering, the weight of the chosen bond and
monomer configuration factorises as
\begin{equation}
 W\left(\{n^b(x)\}, \{m(x)\}\right) = \prod_x \frac{w^{n^b(x)}}{n^b(x)!}
Q(N(x),m(x)) \, .
\end{equation}
Depending on the specific form of the superpotential $P(\phi)$ the
site weight $Q$ might vanish for certain values of $N$ and $m$. This
essentially induces a local constraint on the number of bosonic bonds
attached to a site, e.g.~$N \, \mathrm{mod} \, 2 = 0$ for potentials
even in $\phi$, similar to the constraint on the fermionic bond
occupation numbers. The constraint simply reflects the symmetry
property of the underlying bosonic field and has important
consequences e.g.~for the observables as discussed in section
\ref{sec:observables}.

Let us now consider how the bosonic hopping expansion is modified when
the action with a counterterm, eq.(\ref{eq:action_counterterm}), or
the $Q$-exact action in eq.(\ref{eq:action_Q1}) is employed. While the
counterterm simply changes $V(\phi) \rightarrow V(\phi) + P^{\prime
  \prime}(\phi)/2$ and hence the site weight $Q$, the $Q$-exact
discretisation has a more severe impact on the hopping expansion. To
be more specific, the $Q$-exact actions demand for additional kinds of
bosonic bonds as can be seen by explicitly calculating the term $O$ in
eq.(\ref{eq:Qexact_term}).  Using for example the superpotential $P_u$
we have
\begin{equation}
 O = \sum_x P_u^\prime(\phi_x)(\nabla^- \phi_x) = \sum_x \left\{\mu
   \phi_x^2 + g\phi_x^4 - \mu \phi_x \phi_{x - 1} - g \phi_x^3 \phi_{x
     - 1} \right\}.
\end{equation}
While the first two terms $\mu \phi_x^2$ and $g \phi_x^4$ just modify
the potential $V(\phi)$ describing the local bosonic self-interaction,
the third term $-\mu \phi_x \phi_{x - 1}$ matches the standard hopping
term and modifies the hopping weight $w=1 \, \rightarrow \, w = 1 +
\mu$.  The fourth term $-g \phi_x^3 \phi_{x - 1}$, however, introduces a
new kind of bosonic hopping and hence a new bosonic bond with weight
$g$. Since the hopping carries one power of the bosonic variable
$\phi$ at the left ending and three powers $\phi^3$ at the right
ending the new bosonic bond is directed. In order to distinguish the
two different types of bosonic bonds, we label them by indicating the
number of bosonic variables they carry at each ending, i.e.~$b^B
\rightarrow b^B_{1 \rightarrow 1}$ and $b^B_{1 \rightarrow 3}$ for the
new bond. Of course the new bosonic bond also contributes to the site
occupation number,
\begin{equation}\label{k_Q_1_ex_u}
 N(x) = n^b_{1 \rightarrow 1}(x) + n^b_{1 \rightarrow 1}(x - 1) +
 n^b_{1 \rightarrow 3}(x) + 3 \cdot n^b_{1 \rightarrow 3}(x - 1) \, ,
\end{equation}
and the total weight of a bond configuration becomes
\begin{equation}\label{HT_partition_Q1}
 W =  \prod_x \frac{(1 + \mu)^{n^b_{1 \rightarrow 1}(x)}}{n^b_{1
     \rightarrow 1}(x)!} \frac{g^{n^b_{1 \rightarrow 3}(x)}}{n^b_{1
     \rightarrow 3}(x)!}  Q\left(N(x),m(x) \right) .
\end{equation}

\begin{figure}[thb]
\centering
\includegraphics[width=0.50\textwidth]{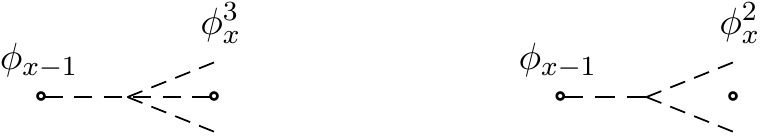}
 \caption{Graphical representation of the bosonic bonds $b^B_{1
     \rightarrow 3}$ and $b^B_{1 \rightarrow 2}$ appearing in the bond formulation for the $Q$-exact action with the superpotentials $P_u$ and $P_b$, respectively.}
  \label{fig:bos_multiple_bonds}
\end{figure}

For the superpotential $P_b$, the explicit expression for the surface
term reads
\begin{equation}
 O = \sum_x P_b^\prime(\phi_x)(\nabla^- \phi_x) = \sum_x \left\{\lambda
     \phi_x^3 - \lambda \phi_x^2 \phi_{x - 1} \right\}.
\end{equation}
The first term $\lambda \phi_x^3$ modifies the local potential
$V(\phi)$ and therefore just changes the site weight $Q$. In contrast
to the previous case there is no additional term $\propto \phi_{x}
\phi_{x-1}$, hence the corresponding hopping weight $w=1$ is
unchanged.  The hopping term $-\lambda \phi_x^2 \phi_{x - 1}$
generates a new type of bosonic bond $b^B_{1 \rightarrow 2}$ with
weight $\lambda$. This directed bond carries one power of the bosonic
variable $\phi$ at the left ending and two powers $\phi^2$ at the
right ending, so the site occupation number is therefore modified as
\begin{equation}\label{k_Q_1_ex_b}
 N(x) = n^b_{1 \rightarrow 1}(x) + n^b_{1 \rightarrow 1}(x - 1) +
 n^b_{1 \rightarrow 2}(x) + 2\cdot n^b_{1 \rightarrow 2}(x - 1) \, .
\end{equation}
Eventually, the total weight of a bond configuration is then found to
be
\begin{equation}\label{HT_partition_Q1_b}
 W = \prod_x \frac{w^{n^b_{1 \rightarrow 1}(x)}}{n^b_{1 \rightarrow
     1}(x)!} \frac{\lambda^{n^b_{1 \rightarrow 2}(x)}}{n^b_{1
     \rightarrow 2}(x)!} Q\left(N(x),m(x) \right) 
\end{equation}
with $w=1$.  In analogy to the illustration for the $b^B_{1
  \rightarrow 1}$ bond in figure \ref{fig:bosonicbond}, we give a
graphical representation of the new bonds $b^B_{1 \rightarrow 3}$ and
$b^B_{1 \rightarrow 2}$ in figure \ref{fig:bos_multiple_bonds}
illustrating their contributions to the site weights at each ends.
As a side remark we note that it is in fact not too surprising to find
directed bosonic hopping terms for the $Q$-exact actions: since these
preserve part of the supersymmetry the oriented fermion hopping needs
to be matched in some way by corresponding oriented boson hopping
terms.

It is straightforward to generalise the above construction to even
more complicated discretisations. For example, we mentioned before
that the addition of the surface term in eq.(\ref{eq:Qexact_term}) to
the original action with any weight different from zero yields a whole
class of $Q$-exact actions \cite{Catterall:2003wd}. Another example is
the discretisation of the action using the Stratanovich prescription
\cite{Ezawa:1985, Beccaria:1998vi, Bergner:2007pu}.  In general, in
addition to the bonds of type $(1 \rightarrow 1)$ and $(1 \rightarrow
2)$ or $(1 \rightarrow 3)$, these actions will also generate bonds of
type $(2 \rightarrow 1)$ or $(3 \rightarrow 1)$ for the
superpotentials $P_b$ and $P_u$. Superpotentials of higher order
produce bonds of correspondingly higher order. All these bonds can be
treated in exactly the same way as discussed above. Each new hopping
of type $(i \rightarrow j)$ induces a new bond $b_{i \rightarrow j}^B$
carrying weight $w_{i \rightarrow j}\equiv w$ and a corresponding bond
occupation number $n^b_{i \rightarrow j}\equiv n$, contributing a
factor $w^{n}/n!$ to the local weight and eventually also modifies the
site occupation number $N$.

\subsection{Partition functions in the loop formulation}
\label{sec:boundary conditions}
\begin{figure}
 \centering
\includegraphics[width=0.72\textwidth]{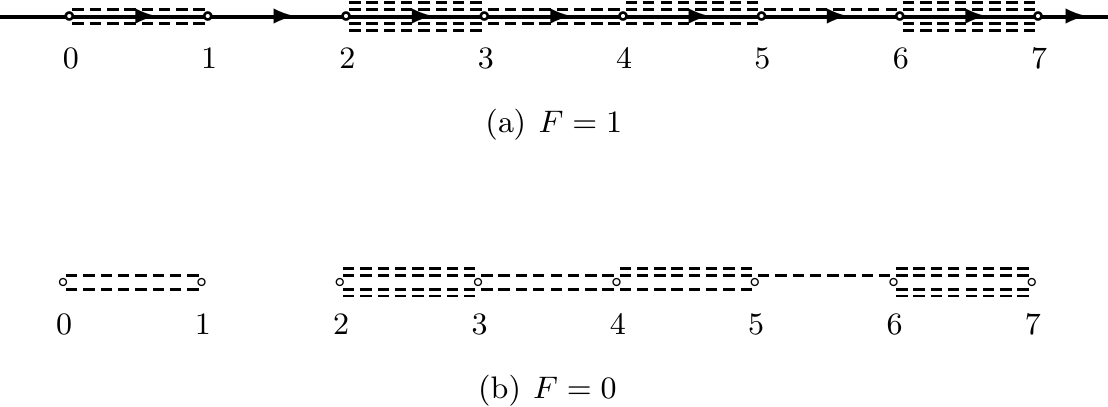}

  \caption{Illustration of a possible bosonic bond configuration in the fermionic sector $F = 1$ and the same configuration in the bosonic sector $F = 0$ on a $L_t = 8$ lattice.}
  \label{illustration_bond_conf}
\end{figure}

After having integrated out the fermionic and bosonic fields $\psibar,
\psi$ and $\phi$, respectively, we are left with discrete fermionic
and bosonic bond occupation numbers as the degrees of freedom. The
path integral has eventually been replaced by a sum over all allowed
bond configurations, possibly restricted by local constraints, and
hence represents a discrete statistical system. By itself this is
already a huge reduction in complexity. Any bond configuration
contributing to the partition function consists of the superposition
of a generic bosonic bond configuration with one of the two allowed
fermionic bond configurations, namely the one representing a closed
fermion loop winding around the lattice or the one without any
fermionic bonds. Therefore, each bond configuration is either
associated with the fermionic sector with fermion number $F=1$, or
with the bosonic sector with $F=0$. In figure
\ref{illustration_bond_conf} we illustrate two such possible
configurations in the fermionic and bosonic sectors on a $L_t = 8$
lattice.
Collecting our results from the previous two sections we can now write
down the contribution of a generic bond configuration
$\C=\{n^b_i(x),m(x)\}$ to the partition function. It depends on the
fermion number and reads
\begin{equation}\label{HT_partition}
 W_F(\C) = \prod_x \left( \prod_i \frac{w_i^{n^b_i(x)}}{n^b_i(x)!} \right)
\prod_x  Q_F(N(x)) 
\end{equation}
where the index $i$ runs over all the types of bosonic bonds appearing
for the specific discretisation under consideration, i.e.~$i\in
\{1\rightarrow 1, 1\rightarrow 2, 1\rightarrow 3\}$. In appendix
\ref{app:discretisations} we summarise the various bond types and
corresponding weights for the discretisations and superpotentials
discussed in the previous two sections. The site weights are given by
\begin{equation}\label{q_f}
 Q_F(N(x)) = \int_{-\infty}^\infty d \phi \ \phi^{N(x)} \e^{-V(\phi)}
 M(\phi)^{1 - F}\, ,
\end{equation}
where the site occupation number $N(x)$ counts all the bosonic bonds
connected to the site $x$,
\begin{equation}
N(x) = \sum_{j,k} \left(j\cdot n^b_{j \rightarrow k}(x) + k\cdot n^b_{j
    \rightarrow k}(x-1) \right) \, .
\end{equation}
The potential $V(\phi)$ depends on the first derivative of the
superpotential, $P^\prime(\phi)$, while the monomer term
$M(\phi)$ depends on second derivative $P^{\prime\prime}(\phi)$ and is
present if the fermion is not ($F=0$) and vice versa ($F=1$). For
superpotentials of polynomial form they can be written as
\begin{equation}
V(\phi) = \sum_n k_n \phi^n \, , \quad M(\phi) = \sum_n m_n \phi^n \, .
\end{equation}
The values of the various coefficients for the superpotentials
discussed in this paper are compiled in the tables in appendix
\ref{app:discretisations}, where we summarise the details of the
various discretisations.  Finally, the full partition functions in the
two sectors can be written as the sum over all configurations $\C$ in
the corresponding configuration space $\Z_F$,
\begin{equation}
 Z_F = \sum_{\C \subset \Z_F} W_{F}(\C) \, .
\end{equation}

The separation of the bond configuration space into the bosonic and
fermionic sectors comes about naturally in the loop formulation, since
the bond configurations fall into separate equivalence classes $\Z_F$
specified by the fermion number $F$.  In principle one can consider
each sector separately and the partition functions simply describe
canonical quantum mechanical systems with fixed fermion number $F=0$
or 1. In terms of a winding fermion this corresponds to boundary
conditions which fix the topology of the winding fermion string, i.e.,
topological boundary conditions.  In order to specify the usual
fermion boundary conditions,
\begin{equation}
 \psibar_{x + L_t} = (-1)^\varepsilon \psibar_x, \qquad \psi_{x + L_t} = (-1)^\varepsilon \psi_x
\end{equation}
with $\varepsilon=0$ and 1 for periodic and antiperiodic boundary
conditions, respectively, the two partition functions need to be
combined. From our discussion in section \ref{subsec:fermion loop
  formulation} we know that the configurations in the fermion sector,
apart from having different weights, pick up a relative sign $(-1)$
coming from the closed fermion loop. An additional sign stems from the
fermion loop crossing the boundary if antiperiodic boundary conditions
are employed. The relative sign between the contributions of the two
sectors can therefore be summarised as $(-1)^{\varepsilon \cdot F}$,
and the partition functions for the systems with periodic or
antiperiodic fermionic boundary conditions can be written as
\begin{equation}\label{Z_PBC}
 Z_{p,a} = Z_{0} \mp Z_{1} \, .
\end{equation}

Depending on the relative size of $Z_0$ and $Z_1$ the combination for
$Z_p$ vanishes or can even be negative. This has important
consequences for the Witten index $W$ which is proportional to
$Z_p$. The index vanishes whenever $Z_0 =Z_1$, i.e., when the
contributions from the bosonic and fermionic sectors cancel each other
exactly. In this case, the free energies of the bosonic and fermionic
vacuum must be equal and hence there exists a gapless, fermionic
excitation which oscillates between the two vacua, namely the
Goldstino mode. As discussed before, the Witten index is regulated at
finite lattice spacing, essentially through the fact that $Z_0$ and
$Z_1$ have different lattice artefacts and therefore do not cancel
exactly. More precisely, the finite lattice spacing breaks the
degeneracy between the vacuum states by inducing a small free energy
difference between the bosonic and fermionic vacua. Consequently, the
Goldstino mode receives a small mass, which only disappears in the
continuum limit, and is hence also regulated. From that point of view
standard Monte Carlo simulations seem to be safe in the sense that
there is no need to simulate at vanishing fermion mass. Nevertheless,
sufficiently close to the massless limit in a supersymmetry broken
system, standard simulation algorithms will almost certainly suffer
from critical slowing down and from fluctuating signs of the
determinant due to the sign problem discussed before.

The separation of the partition function into a bosonic and fermionic
part offers several ways to approach and in fact solve the sign
problem when the supersymmetry is broken. Firstly, one can in
principle perform simulations in each sector separately, but of course
one then misses the physics of the Goldstino mode. Secondly, one can
devise an algorithm which efficiently estimates the relative weights
of the sectors and hence directly probes the signal on top of the
potentially huge cancellations between $Z_0$ and $Z_1$. Fortunately,
such an algorithm is available
\cite{Wenger:2008tq,Wenger:2009mi}. Since this so-called open fermion
string algorithm directly samples the Goldstino mode, there is no
critical slowing down and the physics of the Goldstino is properly
captured. The application of the algorithm to the quantum mechanical
system is the topic of our third paper in the series
\cite{Baumgartner:2015zna}.

Finally, we note that the equivalence classes $\Z_F$ of the bond
configurations specified by the fermion number $F$ can also be
characterised by the winding of the fermion around the lattice.  In
our quantum mechanical system the two characterisations are
equivalent, but in more complicated systems the classification in
terms of the topology of the fermion winding is more appropriate.  It
turns out that the discussion of the topological sectors with fixed
fermion winding number is in fact crucial for the successful
operational application of the fermion loop formulation in more
complicated quantum mechanical systems \cite{Steinhauer:2014oda}, or
in higher dimensions \cite{Wolff:2007ip,Wenger:2008tq}. As a matter of
fact, the separation of the bond configurations into topological
classes provides the basis for the solution of the fermion sign
problem in the ${\cal N}=1$ Wess-Zumino model
\cite{Baumgartner:2011cm,Baumgartner:2011jw,Baumgartner:2013ara,Steinhauer:2014yaa}
in complete analogy to how it is illustrated here in the quantum
mechanical system.

\section{Observables in the loop formulation}
\label{sec:observables}
We now discuss how bosonic and fermionic observables are expressed in
the loop formulation and how the calculation of vacuum expectation
values is affected by the decomposition of the partition function into
its bosonic and fermionic parts.  In general, the expectation value of
an observable $\langle \mathcal{O} \rangle$ is given by
\begin{equation}
 \langle \mathcal{O} \rangle = \frac{1}{Z} \int \mathcal{D} \phi \mathcal{D} \psibar \mathcal{D} \psi \ \mathcal{O}(\phi,\psibar,\psi) \ \e^{-S (\phi,\psibar,\psi)}
\end{equation}
and the explicit expression for periodic and antiperiodic boundary
conditions is
\begin{equation}\label{eq:exp_val}
 \langle \mathcal{O} \rangle_{p,a} = \frac{\langle \! \langle
   \mathcal{O} \rangle \! \rangle_0 \mp \langle \! \langle \mathcal{O}
   \rangle \! \rangle_1}{Z_0 \mp Z_1} \, .
\end{equation}
Here, we have denoted the non-normalised expectation value of the
observable in the sector $F$ by $\langle \! \langle \mathcal{O}
\rangle \! \rangle_F \equiv \langle \mathcal{O} \rangle_F \cdot
Z_F$. According to our discussion at the end of the previous section,
it is important that in order to calculate the expectation values it
is not sufficient to determine $\langle \mathcal{O} \rangle$ in each
sector separately, but it is mandatory to calculate the ratio
$Z_0/Z_1$, or similar ratios which contain the same information such
as $Z_F/(Z_0+Z_1)$.

Recalling that for broken supersymmetry the Witten index is $W=0$, and
hence $Z_{p} = Z_0 -Z_1 = 0$, it is obvious from eq.(\ref{eq:exp_val})
that the vacuum expectation values for periodic boundary conditions
$\langle \O \rangle_{p}$ require a division by zero. Of course this is
simply a manifestation of the fermion sign problem discussed earlier
in section \ref{subsec:fermion_sign_problem}. One might then wonder
whether vacuum expectation values of observables are well defined at
all when the supersymmetry is broken. It turns out, however, that the
finite lattice spacing in fact provides a regularisation for this
problem. For the standard discretisation, supersymmetry is explicitly
broken, such that $Z_{p} \neq 0$ for $a \neq 0$. It is therefore
possible to calculate expectation values for periodic b.c.~at finite
lattice spacing, when they are well defined, and then take the
continuum limit. Whether or not eq.(\ref{eq:exp_val}) with periodic
b.c.~remains finite or diverges in that limit depends on the
observable under consideration. For sensible observables, both the
numerator and the denominator go to zero such that their ratio remains
finite It is then possible to give continuum values for periodic
b.c.~even when the supersymmetry is broken in the continuum and
$Z_0-Z_1 \rightarrow 0$. Sensible observables are those which couple
to the Goldstino mode in the same way as $Z_0-Z_1$ does, i.e.,
observables for which the expectation values in both the bosonic and
fermionic sector converge to the same value towards the continuum
limit. For $Q$-exact discretisations, the situation is more
complicated since in systems with broken supersymmetry $Z_0-Z_1=0$
even at finite lattice spacing. In that case, the physics of the
Goldstino mode is realised exactly at $a\neq 0$. It is then more
useful to calculate observables separately in the fermionic and
bosonic sectors and to verify that they agree.

Important examples for observables are the moments of the bosonic
field and two-point functions. The latter are typically used to
measure the mass gaps in the particle spectrum by extracting the
energy difference between the excited states and the vacuum state, but
they also play important roles in the determination of Ward
identities. In the following subsections, we will derive the
representation of these observables in the loop formulation. This will
turn out to be very useful also for the exact calculation of two-point
functions and other observables using transfer matrices in the second
paper of this series \cite{Baumgartner:2015qba}, where we discuss a
plethora of results, and for the discussion of the simulation
algorithm in the third paper of this series \cite{Baumgartner:2015zna}.

\subsection{Moments of $\phi$}
The expectation value of the $n$-th moment of the field variable
$\phi$ is defined as
\begin{equation}
 \langle \phi^n \rangle = \langle \phi_{x}^n \rangle = \frac{1}{Z} \int \mathcal{D} \phi \mathcal{D} \psibar \mathcal{D} \psi \ \phi_{x}^n \ \e^{-S}.
\end{equation}
When repeating the reformulation in terms of bosonic and fermionic
bonds for this case, it is easy to see that the bond configurations
contributing to the partition functions $Z_F$ also contribute to
$\langle \phi^n \rangle$.  The only difference lies in the weight of
each configuration which is modified due to the additional fields
$\phi_x^n$ present at site $x$. The additional fields only change the
local weight $Q_F(N(x))$ through a change of the local bosonic site
occupation number at site $x$,
\begin{equation}
 N(x) \rightarrow N(x) + n \, .
\end{equation}
Hence, the non-normalised expectation value reads
\begin{align}\label{eq:moment}
\langle \! \langle \phi^n \rangle \! \rangle_F &= \sum_{\C \subset
  \Z_F} \frac{Q_F(N(x) + n)}{Q_F(N(x))} W_{F}(\C) \\
&= \langle \! \langle \frac{Q_F(N(x) +
  n)}{Q_F(N(x))}\rangle \! \rangle_F 
\end{align}
and $\langle \phi^n \rangle$ then follows directly from
eq.(\ref{eq:exp_val}).

We noted earlier that the symmetry properties of the underlying fields
are reflected in local constraints on the bond occupation numbers
which in turn express themselves in the values of the site weights
$Q_F$. As a consequence, the symmetry properties are then also
promoted to the observables through the weights in
eq.(\ref{eq:moment}).  Considering for example potentials $V(\phi)$
even in $\phi$, such as the one following from $P_u$, one finds the
constraint $N \, \mathrm{mod} \, 2 = 0$ which is realised by all site
weights with an odd occupation number being identically zero,
i.e.~$Q_F(N \, \mathrm{mod} \, 2 =1) = 0$. Consequently, the
contributions to odd moments vanish for all bond configurations,
$\langle \phi^n\rangle = 0, n \, \mathrm{odd}$, because $Q_F(N(x) + n)
= 0$.

\subsection{The bosonic $n$-point correlation function}
\begin{figure}
 \centering
\includegraphics[width=0.72\textwidth]{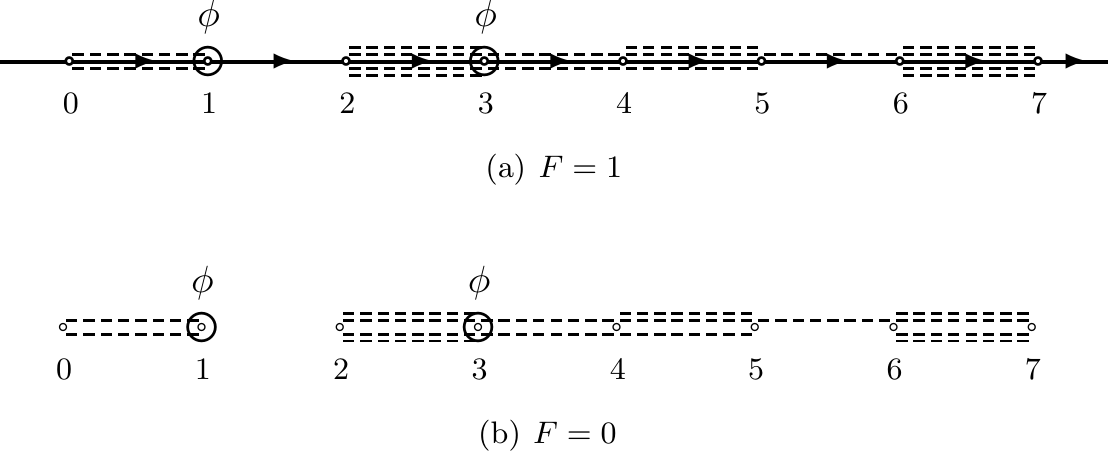}
\caption{Graphical representation of possible configurations similar
  to the closed path configurations in figure
  \ref{illustration_bond_conf}, contributing to the bosonic two-point
  function (a) in the fermionic sector $F = 1$ and (b) the same
  configuration in the bosonic sector $F = 0$ on a $L_t = 8$
  lattice. The additional bosonic variables are marked with a
  $\ocircle$.}
  \label{bos_corr_bond_conf}
\end{figure}

The bosonic two-point function is defined as
\begin{equation}
C^b(x_1-x_2) \equiv \langle \phi_{x_1} \phi_{x_2} \rangle = \frac{g^b(x_1 - x_2)}{Z},
\end{equation}
where
\begin{equation}\label{g_B}
  g^b(x_1 - x_2) = \int \mathcal{D} \phi \mathcal{D} \psibar
  \mathcal{D} \psi \ \phi_{x_1} \phi_{x_2} \ \e^{-S} \equiv \langle \!
  \langle \phi_{x_1} \phi_{x_2} \rangle \! \rangle.
\end{equation}
In the following we will abbreviate the configuration space of the
bosonic two-point function $g^b$ with $\G^b$. It is again
straightforward to rederive the loop formulation in terms of fermionic
and bosonic bond occupation numbers also for this case.  In general
one finds that the bond configurations contributing to $\G^b$ and $\Z$
are the same, but their weights differ due to the insertion of the
additional bosonic field variables $\phi$ at site $x_1$ and $x_2$ in
the configurations contributing to $g^b$. The additional sources only
change the local bosonic site occupation numbers at site $x_1$ and
$x_2$,
\begin{equation}
\label{eq:occupation number 2-point function}
N(x) \rightarrow N(x) + \delta_{x,x_1} + \delta_{x,x_2} \, .
\end{equation}
For $x_1=x_2$ the situation reduces to the one for the second moment
discussed in the previous section.  The fermion number $F$ is not
affected by these sources. Thus, analogously to the configuration
space $\Z$, the configuration space $\G^b$ decomposes into the bosonic
part with $F = 0$ and the fermionic part with $F = 1$. We denote the
separated configuration spaces by adding the subscript $F$, i.e.~$\G^b
\equiv \G^b_F$. In figure \ref{bos_corr_bond_conf} we show two
possible configurations with $F=0$ and $F=1$ contributing to the
bosonic two-point function in the corresponding sectors.
The weight of a configuration where bosonic sources are inserted at
the sites $x_1 \neq x_2$ is then given by
\begin{equation}\label{W_CP_B}
  W_{\G_F^b} = \prod_x \left(\prod_i \frac{w_i^{n_i^b(x)}}{n_i^b(x)!} \right) \left[
    \prod_{x\neq x_1,x_2}  Q_F(N(x)) \right]  Q_F(N(x_1)+1) \cdot  Q_F(N(x_2)+1)
\end{equation}
and the non-normalised expectation value reads
\begin{align}\label{eq:two point function}
\langle \! \langle \phi_{x_1} \phi_{x_2} \rangle \! \rangle_F &= \sum_{\C \subset
  \G^b_F} \frac{Q_F(N(x_1) + 1)}{Q_F(N(x_1))} \cdot
\frac{Q_F(N(x_2) + 1)}{Q_F(N(x_2))} \cdot W_{F}(\C) \\
&= \langle \! \langle \frac{Q_F(N(x_1) +
  1)}{Q_F(N(x_1))} \cdot \frac{Q_F(N(x_2) + 1)}{Q_F(N(x_2))} \rangle \! \rangle_F \, .
\end{align}

It is straightforward to generalise the construction to arbitrary
bosonic $n$-point functions. One simply adds $n$ bosonic sources
$\phi_{x_{k}}^{p_k}, k=1,\ldots,n$ to a given configuration. The
additional sources then contribute to the bosonic site occupation
numbers with additional terms $p_k \cdot \delta_{x,x_{k}}$ modifying
the site weights at positions $x_k$ in analogy to
eq.(\ref{eq:occupation number 2-point function}). Eventually one gets
\begin{equation}\label{eq:n-point function}
\langle \! \langle \phi_{x_1}^{p_1} \ldots \phi_{x_n}^{p_n} \rangle \!
\rangle_F  = 
\langle \! \langle \prod_{k=1}^n \frac{Q_F(N(x_k) +
  p_k)}{Q_F(N(x_k))} \rangle \! \rangle_F \, .
\end{equation}

As discussed before, for some actions there are constraints imposed on
the bond configurations reflecting the symmetry properties of the
bosonic field. In such a case, the bond configurations in the
configuration spaces $\Z_F$ and $\G^b_F$ need no longer be the
same. Considering again the example of a potential $V(\phi)$ even in
$\phi$ such that the parity transformation $\phi \rightarrow -\phi$ is
a symmetry of the action, the constraint $N \, \mathrm{mod} \, 2 = 0$
requires an odd number of bosonic bonds connected to a site containing
an odd power of $\phi$ as a source term, but the corresponding
underlying bond configuration contributes with weight zero to
$\Z_F$. Hence the sets of configurations with nonvanishing weights
contributing to $Z_F$ and $g^b_F$ have no overlap. In addition, from
the symmetry it follows that operators with different quantum numbers,
in this case the parity, do not mix, i.e.~their correlation is exactly
zero, e.g.~$\langle \phi_{x_1}^2 \phi_{x_2}\rangle = 0$. It is easy to
see that this property is strictly enforced in the loop formulation,
since there exist no bond configurations which can accommodate the
sources and fulfil the constraints $N(x) \, \mathrm{mod} \, 2 = 0$ at
the same time.

\subsection{The fermionic correlation function}
The fermionic two-point correlation function is defined as
\begin{equation}
 C^f(x_1 - x_2) \equiv \langle \psi_{x_1} \psibar_{x_2} \rangle = \frac{g^f(x_1 - x_2)}{Z},
\end{equation}
where
\begin{equation}
  g^f(x_1 - x_2) = \int \mathcal{D} \phi \mathcal{D} \psibar
  \mathcal{D} \psi \ \psi_{x_1} \psibar_{x_2} \ \e^{-S} \equiv \langle
  \! \langle  \psi_{x_1} \psibar_{x_2} \rangle \! \rangle \, .
\end{equation}
Similarly to the bosonic correlation function, configurations
contributing to the fermionic correlation function have additional
fermionic variables $\psibar$ and $\psi$ inserted in the path integral
at positions $x_1$ and $x_2$. We will refer to these variables as the
source and the sink, respectively. To derive the weight of a
configuration in the configuration space $\G^f$ of the fermionic
two-point functions, we repeat the expansion of the fermionic
Boltzmann factor in eq.(\ref{expansion_fermionic}) while including the
additional fermionic variables. The expansion yields
\begin{equation}
\begin{split}
 &\psi_{x_1} \psibar_{x_2} \e^{-S_\Lambda^F} = \\
 & \quad \psi_{x_1} \psibar_{x_2} \prod_x \left( \sum_{m(x) = 0}^1   \! \! \Big(-M(\phi_x) \psibar_x \psi_x \Big)^{m(x)} \right) \prod_x \left( \sum_{n^f(x) = 0}^1 \!\!\Big( \psibar_x \psi_{x - 1}\Big)^{n^f(x)} \right)
\end{split}
\end{equation}
and the subsequent Grassmann integration, still requiring exactly one
pair of variables $\psibar$ and $\psi$ at each site $x$, yields an
adjustment of the fermionic occupation numbers $m(x)$ and $n^f(x)$ in
order to obtain a nonvanishing contribution to the two-point function.

\begin{figure}
 \centering

\includegraphics[width=0.72\textwidth]{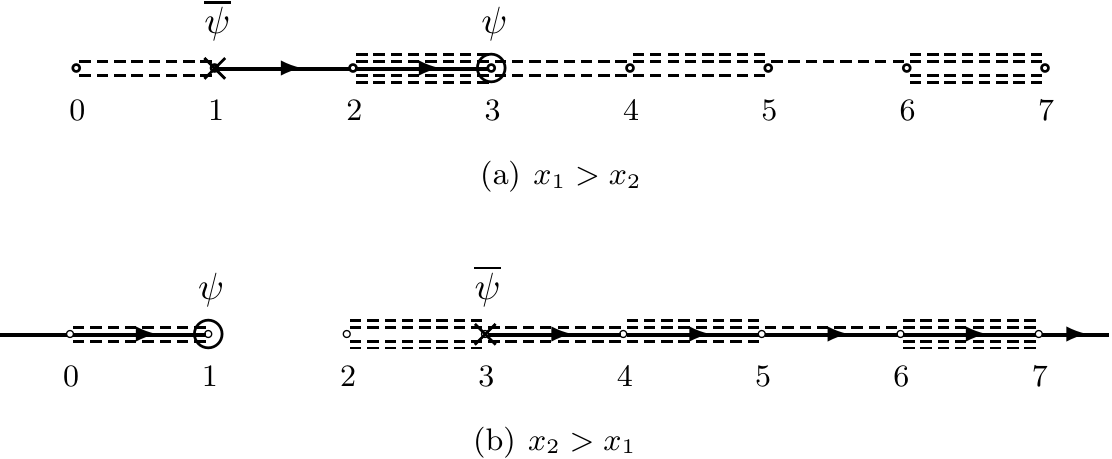}
\caption{Graphical representation of possible configurations similar
  to the constrained path configurations in figure
  \ref{illustration_bond_conf}, contributing to the fermionic
  two-point function (a) for $x_1 > x_2$ and (b) the same
  configuration for $x_2 > x_1$ on a $L_t = 8$ lattice. The additional
  variables are marked with a $\ocircle $ for $\psi_{x_1}$ and a
  $\times$ for $\psibar_{x_2}$.}
  \label{fer_corr_bond_conf}
\end{figure}

We first consider the case where $x_1 = x_2 \equiv y$. It is easy to
see that the only possibility to saturate each site is given by the
choice
\begin{eqnarray}\label{ferm_bond_nr}
 n^f(x) & = & 0 \qquad \forall x, \\
m(x) & = &\left\{ \begin{array}{ll}
0 & \mathrm{if} \ x = y,\\
1 &  \textrm{else}.
\end{array} \right.
\end{eqnarray}
For such a configuration, the site $y$ is saturated through the source
and the sink, yielding a factor 1 as the fermionic contribution to the
bosonic integration. All other sites are saturated via the monomer
terms which have to be accounted for by including the corresponding
factors $M(\phi)$ into the bosonic integration for each of these
sites, so one eventually obtains
\begin{align}
\langle \! \langle \psi_{y} \psibar_y \rangle \!
\rangle_0  
&= \sum_{\C \subset \G^f} \prod_x \left(\prod_i \frac{w_i^{n_i^b(x)}}{n_i^b(x)!} \right)
\left[ \prod_{x\neq y}  Q_0(N(x)) \right] \cdot Q_1(N(y)) \\ 
&= \sum_{\C \subset \G^f} \frac{Q_1(N(y))}{Q_0(N(y))} W_0(\C) \\
& = \langle \! \langle \frac{Q_1(N(y))}{Q_0(N(y))} \rangle \! \rangle_0 \, .
\end{align}
If the additional fermionic variables are not at the same site, $x_1
\neq x_2$, source and sink can only be paired with the ending of a
fermionic bond. Keeping in mind that the fermionic bonds are directed,
it is straightforward to see that one needs $(x_1 - x_2) \
\mathrm{mod} \ L_t$ of these bonds to connect the source with the
sink, thus forming an open fermionic string. This is illustrated in
figure \ref{fer_corr_bond_conf} where we show two typical bond
configurations using the symbols $\ocircle$ and $\times$ to denote the
sink $\psi_{x_1}$ and the source $\psibar_{x_2}$, respectively.
It is clear that each site along the open fermion string is
automatically saturated by the variables of one ingoing and one
outgoing fermionic bond. Those sites and the ones which are saturated
with either the source or the sink and a fermionic bond attached to it
yield a factor 1 as the fermion contribution to the bosonic
integration, while all other sites contribute with the monomer weight
$M(\phi)$.

Because the fermionic bonds are directed, the order of the source and
the sink matters and we need to distinguish between the cases $x_2 >
x_1$ and $x_1 > x_2$. For $x_2 > x_1$, the open string connects source
and sink without crossing the boundary and each configuration is
characterised by the numbers
\begin{eqnarray}
 n^f(x) & = &\left\{ \begin{array}{ll}
1 & \mathrm{if} \ x_2 \leq x < x_1,\\
0 & \textrm{else},
\end{array} \right. \\
m(x) & = &\left\{ \begin{array}{ll}
0 & \mathrm{if} \ x_2 \leq x \leq x_1,\\
1 & \textrm{else},
\end{array} \right. 
\end{eqnarray}
while for $x_1 < x_2$, the fermionic string crosses the boundary and
the numbers to characterise the configuration are given by
\begin{eqnarray}
 n^f(x) & = &\left\{ \begin{array}{ll}
0 & \mathrm{if} \ x_1 \leq x < x_2,\\
1 & \textrm{else},
\end{array} \right. \\
m(x) & = &\left\{ \begin{array}{ll}
1 & \mathrm{if} \ x_1 < x < x_2,\\
0 & \textrm{else}.
\end{array} \right.
\end{eqnarray}
Whether or not the open fermionic string crosses the boundary of the
lattice is relevant for the overall sign of the configuration. Namely,
the crossing yields one extra factor of $(-1)$ for antiperiodic
boundary conditions, and this has to be taken into account in the
overall book keeping for the 2-point function.

\begin{figure}
\centering
$\xymatrix{
& \qquad \G^f \qquad \ar @/^/ [dl]^{\ominus \psibar \psi}  \ar @/^/[dr]^{ \ominus \psibar \psi}& \\
\Z_0 \ar @/^/[ur]^{\oplus \psibar \psi}  \ar @/^/[dd]^{ \oplus \phi \phi} &  & \Z_1 \ar @/^/ [ul]^{\oplus \psibar \psi}  \ar @/^/[dd]^{\oplus \phi \phi}\\
& & \\
\G^b_0 \ar @/^/[uu]^{\ominus \phi \phi} & & \G_1^b \ar @/^/[uu]^{\ominus \phi \phi} }$
\caption{Schematic representation of the configuration spaces. The
  configuration space $\G^f \equiv \G^f_0 = \G^{\overline{f}}_1$
  mediates between the bosonic and the fermionic sector. By the
  symbols $\oplus$ and $\ominus$, we denote the addition and removal
  of the source and sink field variables, respectively.}
  \label{conf_diag}
\end{figure}
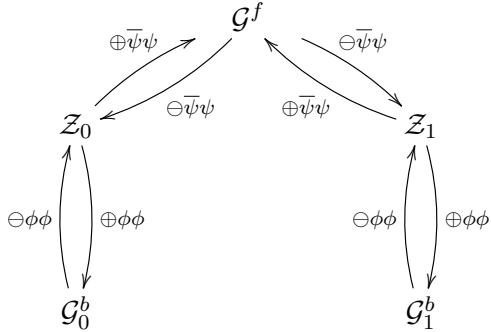

We are now able to give an explicit expression for the weight of an
open fermion string configuration in $\G^f$ contributing to
$C^f(x_1-x_2)$. Each site $x_i$ contributing a factor 1 to the bosonic
integration amounts to a site weight $Q_1(N(x_i))$, while a site $x_j$
contributing the monomer weight $M(\phi_{x_j})$ to the bosonic
integration yields a site weight $Q_0(N(x_j))$.  To simplify the
notation we define the set $\mathcal{F}$ of lattice sites belonging to
the open fermion string as
\begin{eqnarray}
 \mathcal{F}(x_1,x_2) & = &\left\{ \begin{array}{ll}
\{ x \in \Lambda \ | \ x_2 \leq x \leq x_1 \} & \mathrm{if} \ x_2 \leq x_1,\\
\{ x \in \Lambda \ | \ x \leq x_1 \cup x \geq x_2 \} & \mathrm{if} \ x_1 < x_2.
\end{array} \right.
\end{eqnarray}
The weight of a configuration contributing to $\G^f$ can then be
written as
\begin{equation}\label{W_f}
 W_{\G^f} = \prod_x \left( \prod_i \frac{w_i^{n_i^b(x)}}{n_i^b(x)!}
 \right) \left[ \prod_{x \in \phantom{\notin} \! \! \! \! \mathcal{F}}
   Q_1(N(x)) \right] \left[ \prod_{x \notin \mathcal{F}} Q_0(N(x))
 \right] \, ,
\end{equation}
and the non-normalised expectation value of the fermionic two-point
function is
\begin{align}
\langle \! \langle \psi_{x_1} \psibar_{x_2} \rangle \!
\rangle_0  
&= \sum_{\C \subset \G^f} W_{\G^f}(\C) \\
&= \sum_{\C \subset \G^f} \left[\prod_{x \in \phantom{\notin} \! \! \! \!
  \mathcal{F}} \frac{Q_1(N(x))}{Q_0(N(x))} \right] \cdot W_0(\C) \\
& = \langle \! \langle \prod_{x \in \phantom{\notin} \! \! \! \!
  \mathcal{F}} \frac{Q_1(N(x))}{Q_0(N(x))}\rangle \! \rangle_0 \, .
\end{align}
This result implies that the configuration space $\G^f$ does not
decompose into the bosonic and fermionic sector $F = 0$ and $F =
1$. Rather, all configurations in the configuration space of fermionic
two point functions are associated with the bosonic sector. In a way,
the configuration space $\G^f$ mediates between the bosonic and the
fermionic sectors $\Z_0$ and $\Z_1$. The transition from one
configuration space to another is induced by adding or removing the
additional field variables $\psibar \psi$.  The relation between the
various bond configuration spaces is schematically illustrated in
figure \ref{conf_diag}.
The picture suggests to interpret the fermionic correlation function
$C^f(x - y)$ as an open fermion string on the background of bosonic
bond configurations in sector $\Z_0$, or as an open antifermion string
on the background of bond configurations in sector $\Z_1$, i.e., as a
antifermionic correlation function $-C^{\overline{f}}(y - x)$. It is
this property which forms the basis for an efficient simulation
algorithm which will be discussed in detail in the third paper of this
series \cite{Baumgartner:2015zna}.

Finally, the reformulation of the fermionic correlation functions in
terms of bond variables can be generalised to include more complicated
fermionic source and sink operators such as $\psi_x \phi_x^k$ or
$\psibar_x \phi_x^k$. The construction is rather straightforward and
yields
\begin{align}
\langle \! \langle \psi_{x_1} \phi_{x_1}^k \cdot \psibar_{x_2}\phi_{x_2}^l \rangle \!
\rangle_0  
&= \sum_{\C \subset \G^f} \left[\prod_{x \in \phantom{\notin} \! \! \! \!
  \mathcal{F}} \frac{Q_1(N(x)+k\cdot \delta_{x,x_1}+l\cdot \delta_{x,x_2})}{Q_0(N(x))} \right] \cdot W_0(\C) \\
& = \langle \! \langle \prod_{x \in \phantom{\notin} \! \! \! \!
  \mathcal{F}} \frac{Q_1(N(x)+k\cdot \delta_{x,x_1}+l\cdot \delta_{x,x_2})}{Q_0(N(x))}\rangle \! \rangle_0 \, ,
\end{align}
i.e., only the site occupation numbers at site $x_1$ and $x_2$ are
modified accordingly. Similarly to the discussion concerning the
bosonic $n$-point correlation function, operators with different
quantum numbers, for example the parity for actions symmetric under
$\phi \rightarrow - \phi$, do not mix if the symmetry is intact,
e.g.~$\langle \psi_{x_1}\phi_{x_1}^2 \cdot
\psibar_{x_2}\phi_{x_2}\rangle = 0$. It is again easy to see that this
property is strictly enforced through the constraints $N(x) \,
\mathrm{mod} \, 2 = 0$ for parity symmetric actions.

\section{Conclusions}

Simulations of supersymmetric models on the lattice with
(spontaneously) broken supersymmetry suffer from a fermion sign
problem related to the vanishing of the Witten index. This problem is
a generic one and must occur whenever a massless Goldstino mode is
present in the system. In this paper we discussed a novel approach
which solves this problem for ${\cal N}=2$ supersymmetric quantum
mechanics by formulating its Euclidean path integral on the lattice in
terms of fermion loops. The formulation is based on the exact hopping
expansion of the fermionic action and allows the explicit
decomposition of the partition function into a bosonic and a fermionic
sector associated with the corresponding vacua. Since the two vacua
contribute with opposite signs, the separation isolates the cause of
the sign problem and opens the way for its solution. In fact, the
explicit separation of the sectors in the fermion loop formulation
allows the construction of a simulation algorithm which samples these
sectors separately, and more importantly also samples the relative
weights between them. We demonstrate in the third paper of this series
\cite{Baumgartner:2015zna} that in this way, the loop formulation indeed
provides a solution to the fermion sign problem. The solution is not
restricted to the quantum mechanics case, but it is in fact also
applicable in higher dimensions. In particular, it also applies to the
supersymmetric ${\cal N} =1$ Wess-Zumino model
\cite{Steinhauer:2014yaa}, where the formulation has proven to
successfully solve the fermion sign problem.

In addition to the sign problem, in this paper we have discussed
various discretisation schemes for regularising ${\cal N}=2$
supersymmetric quantum mechanics on the lattice using Wilson
fermions. Because the lattice formulations break supersymmetry
explicitly, special care has to be taken to guarantee the restoration
of the supersymmetries in the continuum limit. A very straightforward
discretisation for example requires the addition of a single
counterterm which compensates certain perturbative loop
corrections. We demonstrate explicitly by means of the boson and
fermion mass spectra how the absence of such a term spoils the correct
continuum limit. Another discretisation is based on the insertion of
the Wilson term directly in the superpotential. This construction
leads to the $Q$-exact discretisation which maintains one of the two
supersymmetries exactly at finite lattice spacing. This eventually
guarantees the automatic restoration of the full supersymmetries in
the continuum.

For both discretisation schemes, in addition to the fermion loop
reformulation, we have reformulated the quantum mechanics system on
the lattice in terms of bosonic bonds. As in the fermionic case, the
formulation is based on the exact hopping expansion of the bosonic
actions.  The bosonic bond formulation is not a necessary ingredient
in the solution of the fermion sign problem, but completes the
description of the quantum mechanical system in terms of solely
discrete variables. In fact, while the fermion loop formulation is not
affected by the choice of discretisation, the bond formulation is and
in general requires arbitrary types of bonds beyond the simple one. We
have discussed in detail how the simple bosonic bond formulation needs
to be adapted in order to accommodate more complicated
discretisations, such as the $Q$-exact one, as well as arbitrary
superpotentials.

Furthermore, we also derived explicit expressions for various
observables in the bond formulation, such as moments of the bosonic
field, bosonic $n$-point correlation functions, and fermionic 2-point
functions with arbitrary fermionic operators. For the latter we
emphasised its interpretation as an open fermion string. In addition,
we argued that the fermion correlator in the bosonic vacuum can
equally well be interpreted as the antifermion correlator in the
fermionic vacuum. More importantly, however, is the fact that the
configurations including the open fermion string represent the
configuration space mediating between the bosonic and fermionic
configuration spaces. This eventually forms the basis for the
efficient simulation algorithm discussed in paper three of this series
\cite{Baumgartner:2015zna}.

Finally, as an outlook, we point out that using the bond formulation
it is straightforward to construct transfer matrices separately for
the bosonic and fermionic sector. They allow in turn to solve the
lattice system exactly. This construction and the subsequent solution
will be the subject of the second paper in this series
\cite{Baumgartner:2015qba}.

\begin{appendix}
\section{Summary of the discretisations}\label{app:discretisations}
In this appendix, we write out explicitly the actions for which we
discussed in detail the derivation of the loop formulation.  The
generic lattice actions are given by eqs.(\ref{eq:S_stand_a}),
(\ref{eq:action_counterterm}) and (\ref{eq:action_Q1}),
\begin{eqnarray}
 S_L & = & \sum_x \Big\{\frac{1}{2}(\nabla^-\phi_x)^2 +
   \frac{1}{2}P^\prime(\phi_x)^2 + \overline{\psi}_x(\nabla^- +
   P^{\prime \prime}(\phi_x))\psi_x \Big\} \, ,\\
  S_L^c & = & S_L + \frac{1}{2}\sum_x P^{\prime \prime}(\phi_x) \, ,\\
 S_L^{Q} & = & S_L + \sum_x P^\prime(\phi_x)(\nabla^- \phi_x) \, . 
\end{eqnarray}
For the polynomial superpotentials discussed in this paper the
resulting bosonic self-interaction $V(\phi)$ and the fermionic monomer
term $M(\phi)$ can be described by
\begin{equation}
 V(\phi) = \sum_{n = 1}^6 k_n \phi^n, \qquad M(\phi) = \sum_{n = 0}^2 m_n \phi^n.
\end{equation}
The weights of the directed bosonic bonds are given by $w_{1
  \rightarrow n}$, where $n$ indicates the number of bosonic sources
carried at the right ending of the specific bond. In the following we
explicitly write out the actions and tabulate the coefficients $k_n$
and $m_n$ as well as the bond weights $w_{1 \rightarrow n}$ for the
two superpotentials
\begin{eqnarray}
 P_u(\phi) & = & \frac{1}{2}\mu \phi^2 + \frac{1}{4}g \phi^4,\\
 P_b(\phi) & = & -\frac{\mu^2}{4 \lambda} \phi + \frac{1}{3}\lambda \phi^3,
\end{eqnarray}
which yield systems with unbroken and broken supersymmetry,
respectively.

\subsection{The actions for the superpotential $P_u$}
Writing out explicitly the actions for the superpotential $P_u$, we
have
\begin{eqnarray}\nonumber
S_L &  =  & \sum_x \Big\{ \frac{1}{2} \left(2 + \mu^2 \right)\phi_x^2 + \mu g\phi_x^4 + \frac{1}{2}g^2\phi_x^6 - \phi_x\phi_{x-1} \\
& & \qquad  + \left(1 + \mu + 3g\phi_x^2 \right)\overline{\psi}_x\psi_x - \overline{\psi}_x\psi_{x-1} \Big\}, \\ \nonumber
S_L^c &  =  & \sum_x \Big\{ \frac{1}{2}\left(2 + \mu^2 + 3g \right)\phi_x^2 + \mu g\phi_x^4 + \frac{1}{2}g^2\phi_x^6 - \phi_x\phi_{x-1}\\
&  & \qquad + \left(1 + \mu + 3g\phi_x^2 \right)\overline{\psi}_x\psi_x - \overline{\psi}_x\psi_{x-1} \Big\},\\ \nonumber
S_L^{Q} &  = & \sum_x \Big\{\frac{1}{2}\left(2 + 2\mu + \mu^2 \right)\phi_x^2 + g(\mu + 1)\phi_x^4 + \frac{1}{2}g^2\phi_x^6 - g\phi_x^3\phi_{x-1}\\ 
&  & \qquad - \left(1 + \mu \right)\phi_x\phi_{x-1} + \left(1 + \mu +
  3g\phi_x^2 \right)\overline{\psi}_x\psi_x -
\overline{\psi}_x\psi_{x-1}\Big\}, 
\end{eqnarray}
and the coefficients and hopping weights can directly be read
off. They are compiled in table \ref{tab:coefficients Pu}.
\begin{table}[htp!]
 \centering
\begin{tabular}{c c c c}
\toprule
  & $S_L$ & $S_L^c$& $S_L^{Q}$  \\
\midrule
 $k_1$ & 0 & 0 & 0  \\
 $k_2$ & $1 + \frac{1}{2}\mu^2$ & $1 + \frac{1}{2}\mu^2 + \frac{3}{2}g$ & $1 + \mu + \frac{1}{2}\mu^2$  \\
 $k_3$ & 0 & 0 & 0  \\
 $k_4$ & $\mu g$ & $\mu g$ & $g(1 + \mu)$  \\
 $k_5$ & 0 & 0 & 0  \\
 $k_6$ & $\frac{1}{2}g^2$ & $\frac{1}{2}g^2$ & $\frac{1}{2}g^2$  \\
 $w_{1 \rightarrow 1}$ & $1$ & $1$ & $1 + \mu$  \\
 $w_{1 \rightarrow 3}$ & $0$ & $0$ & $g$  \\
 $m_0$ & $1 + \mu$ & $1 + \mu$ & $1 + \mu$  \\
 $m_1$ & $0$ & $0$ & $0$  \\
 $m_2$ & $3g$ & $3g$ & $3g$  \\
\bottomrule
\end{tabular}
\caption{Unbroken supersymmetric quantum mechanics: coefficients and
  hopping weights for the superpotential $P_u(\phi) = \frac{1}{2}\mu
  \phi^2 + \frac{1}{4}g \phi^4$.
\label{tab:coefficients Pu}
}
\end{table}

\newpage
\subsection{The actions for the superpotential $P_b$}
So far we have concentrated on the superpotential with broken
supersymmetry of the form
\begin{equation}
\label{eq:unshifted superpotential}
P_b(\phi) = -\frac{\mu^2}{4\lambda} \phi + \frac{1}{3} \lambda\phi^3, 
\end{equation}
yielding a potential for the bosonic field which is symmetric under
the parity transformation $\phi \rightarrow -\phi$. Writing out
explicitly the actions for this superpotential we obtain
\begin{eqnarray}\nonumber
S_L &  =  & \sum_x \Big\{ \frac{1}{2} \left(2 - \frac{\mu^2}{2} \right)\phi_x^2 + \frac{1}{2}\lambda^2\phi_x^4 - \phi_x\phi_{x-1} \\
 & & \qquad  + \left(1 + 2\lambda \phi_x^2 \right)\overline{\psi}_x\psi_x - \overline{\psi}_x\psi_{x-1} \Big\}, \\ \nonumber
S_L^c &  =  & \sum_x \Big\{ \lambda \phi_x + \frac{1}{2}\left(2 - \frac{\mu^2}{2}\right)\phi_x^2 + \frac{1}{2}\lambda^2\phi_x^4 - \phi_x\phi_{x-1}\\
&  & \qquad + \left(1 + 2\lambda \phi_x^2 \right)\overline{\psi}_x\psi_x - \overline{\psi}_x\psi_{x-1} \Big\},\\ \nonumber
S_L^{Q} &  = & \sum_x \Big\{\frac{1}{2}\left(2 - \frac{\mu^2}{2}
\right)\phi_x^2 + \lambda \phi_x^3 + \frac{1}{2}\lambda^2\phi_x^4 - \lambda \phi_x^2\phi_{x-1}\\ 
&  & \qquad - \phi_x\phi_{x-1} + \left(1 + 2 \lambda \phi_x^2 \right)\overline{\psi}_x\psi_x - \overline{\psi}_x\psi_{x-1}\Big\},
\end{eqnarray}
and the corresponding coefficients and hopping weights are given in
table \ref{tab:coefficients unshifted Pb}.
\begin{table}[htp!]
 \centering
\begin{tabular}{c c c c}
\toprule
  & $S_L$ & $S_L^c$& $S_L^{Q}$  \\
\midrule
 $k_1$ & 0 & $\lambda$ & 0  \\
 $k_2$ & $\frac{1}{4}(4 - \mu^2)$ & $\frac{1}{4}(4 - \mu^2)$ & $\frac{1}{4}(4 - \mu^2)$  \\
 $k_3$ & 0 & 0 & $\lambda$  \\
 $k_4$ & $\frac{1}{2} \lambda^2$ & $\frac{1}{2} \lambda^2$ & $\frac{1}{2} \lambda^2$  \\
 $k_5$ & 0 & 0 & 0  \\
 $k_6$ & 0 & 0 & 0  \\
 $w_{1 \rightarrow 1}$ & $1$ & $1$ & $1$  \\
 $w_{1 \rightarrow 2}$ & 0 & 0 & $\lambda$  \\
 $m_0$ & $1$ & $1$ & $1$  \\
 $m_1$ & $2 \lambda$ & $2 \lambda$ & $2 \lambda$  \\
 $m_2$ & $0$ & $0$ & $0$  \\
\bottomrule
\end{tabular}
\caption{Broken supersymmetric quantum mechanics: coefficients and
  hopping  weights for the superpotential $P_b(\phi) = -\frac{\mu^2}{4
    \lambda}\phi + \frac{1}{3} \lambda \phi^3$.
\label{tab:coefficients unshifted Pb}
}
\end{table}
\newpage

In order to apply perturbation theory it is more useful to consider
the shifted superpotential
\begin{equation}
P_b(\phi) = \frac{1}{2} \mu \phi^2 + \frac{1}{3} \lambda \phi^3
\end{equation}
which is obtained from eq.(\ref{eq:unshifted superpotential}) by
applying the shift $\phi \rightarrow \phi +\mu/2\lambda$ and
neglecting any constant terms. The potential for the bosonic field
then has a minimum at $\phi=0$, but the parity symmetry is no longer
manifest.  Writing out explicitly the actions for this form of the
superpotential $P_b$, we have
\begin{eqnarray}\nonumber
S_L &  =  & \sum_x \Big\{ \frac{1}{2} \left(2 + \mu^2 \right)\phi_x^2 + \mu \lambda\phi_x^3 + \frac{1}{2}\lambda^2\phi_x^4 - \phi_x\phi_{x-1} \\
 & & \qquad  + \left(1 + \mu + 2\lambda \phi_x^2 \right)\overline{\psi}_x\psi_x - \overline{\psi}_x\psi_{x-1} \Big\}, \\ \nonumber
S_L^c &  =  & \sum_x \Big\{ \lambda \phi_x + \frac{1}{2}\left(2 + \mu^2\right)\phi_x^2 + \mu \lambda \phi_x^3 + \frac{1}{2}\lambda^2\phi_x^4 - \phi_x\phi_{x-1}\\
&  & \qquad + \left(1 + \mu + 2\lambda \phi_x^2 \right)\overline{\psi}_x\psi_x - \overline{\psi}_x\psi_{x-1} \Big\},\\ \nonumber
S_L^{Q} &  = & \sum_x \Big\{\frac{1}{2}\left(2 + 2\mu + \mu^2 \right)\phi_x^2 + \lambda(\mu + 1)\phi_x^3 + \frac{1}{2}\lambda^2\phi_x^4 - \lambda \phi_x^2\phi_{x-1}\\ 
&  & \qquad - \left(1 + \mu \right)\phi_x\phi_{x-1} + \left(1 + \mu + 2 \lambda \phi_x^2 \right)\overline{\psi}_x\psi_x - \overline{\psi}_x\psi_{x-1}\Big\},
\end{eqnarray}
and the corresponding coefficients and hopping weights are given in
table \ref{tab:coefficients Pb}.
\begin{table}[htp!]
 \centering
\begin{tabular}{c c c c}
\toprule
  & $S_L$ & $S_L^c$& $S_L^{Q}$  \\
\midrule
 $k_1$ & 0 & $\lambda$ & 0  \\
 $k_2$ & $\frac{1}{2}(2 + \mu^2)$ & $\frac{1}{2}(2 + \mu^2)$ &
 $\frac{1}{2}(2 + 2\mu + \mu^2)$  \\
 $k_3$ & $\mu \lambda$ & $\mu \lambda$ & $(1+\mu)\lambda$  \\
 $k_4$ & $\frac{1}{2} \lambda^2$ & $\frac{1}{2} \lambda^2$ & $\frac{1}{2} \lambda^2$  \\
 $k_5$ & 0 & 0 & 0  \\
 $k_6$ & 0 & 0 & 0  \\
 $w_{1 \rightarrow 1}$ & $1$ & $1$ & $1+\mu$  \\
 $w_{1 \rightarrow 2}$ & 0 & 0 & $\lambda$  \\
 $m_0$ & $1 + \mu$ & $1 + \mu$ & $1 + \mu$  \\
 $m_1$ & $2 \lambda$ & $2 \lambda$ & $2 \lambda$  \\
 $m_2$ & $0$ & $0$ & $0$  \\
\bottomrule
\end{tabular}
\caption{Broken supersymmetric quantum mechanics: coefficients and
  hopping weights for the superpotential $P_b(\phi) = \frac{1}{2}\mu\phi^2 + \frac{1}{3} \lambda \phi^3$.}
\label{tab:coefficients Pb}
\end{table}

\end{appendix}

\bibliography{susyQM_LoopFormulation}
\bibliographystyle{JHEP}
\end{document}